\documentclass[reprint,superscriptaddress,amsmath,amssymb,aps]{revtex4-2}

\usepackage{graphicx}
\usepackage{xcolor}
\usepackage{dcolumn}
\usepackage{bm}
\usepackage{amsthm}
\usepackage{algpseudocode}

\usepackage{braket}
\usepackage{lipsum}
\usepackage{tikz}  
\usepackage{ctable}
\usepackage{tabularx}
\usepackage{array}
\usepackage{float}
\usepackage{graphicx}
\usepackage[colorlinks = true,
            linkcolor = blue,
            urlcolor  = black,
            citecolor = blue,
            anchorcolor = blue]{hyperref}
 
\usepackage{cleveref}
\usepackage[markup=underlined,draft]{changes}
\definechangesauthor[name=Gerhard, color=magenta]{GH}
\definechangesauthor[name=Martin, color=red]{MZ}
\definechangesauthor[name=Martin, color=orange]{VD}

\begin{document}

\title{Extrapolation method to optimize linear-ramp quantum approximate optimization algorithm parameters: Evaluation of runtime scaling}

\author{Vanessa Dehn\thanks{Fraunhofer Institute for Applied Solid State Physics IAF}}
\affiliation{%
 Fraunhofer Institute for Applied Solid State Physics IAF, Tullastr. 72, 79108 Freiburg, Germany
}%
\author{Martin Zaefferer}
\affiliation{%
DHBW Ravensburg, Marienplatz 2, 88212 Ravensburg, Germany
}%
\author{Gerhard Hellstern}
\affiliation{%
DHBW Stuttgart, Herdweg 20, 70174 Stuttgart, Germany
}%
\author{Karthik Jayadevan}
\affiliation{%
 Fraunhofer Institute for Applied Solid State Physics IAF, Tullastr. 72, 79108 Freiburg, Germany
}%
\author{Florentin Reiter}
\affiliation{%
 Fraunhofer Institute for Applied Solid State Physics IAF, Tullastr. 72, 79108 Freiburg, Germany
}%
\author{Thomas Wellens}%
 \email{thomas.wellens@iaf.fraunhofer.de}
\affiliation{%
 Fraunhofer Institute for Applied Solid State Physics IAF, Tullastr. 72, 79108 Freiburg, Germany
}%

\begin{abstract}
\noindent The Quantum Approximate Optimization Algorithm (QAOA) 
has been suggested as a promising candidate for the solution of
combinatorial optimization problems. Yet, whether - or under what conditions - it may offer an advantage compared to classical algorithms remains to be proven. 
Using the standard variational form of QAOA requires a high number of circuit parameters that have
to be optimized at a sufficiently large depth, which constitutes a bottleneck for achieving a potential scaling advantage. 
The linear-ramp QAOA (LR-QAOA) has been proposed to address this issue, as it relies on only two parameters which have to be optimized. Based on this, we develop a method to estimate suitable values for those parameters through extrapolation, starting from smaller problem sizes (number of qubits) towards larger problem sizes. 
We apply this method to several use cases such as portfolio optimization, feature selection, clustering 
and weighted maxcut. From results obtained on a noiseless quantum emulator, we evaluate
the quantum runtime scaling for finding the optimal solution and compare it with that of classical methods. In the case of portfolio optimization, we demonstrate superior scaling compared to the classical runtime
for the problem sizes of up to $28$ qubits that we consider  in this work.
\end{abstract}

\maketitle

\section{\label{sec:intro}Introduction}

The quantum approximate optimization algorithm (QAOA) is a widely studied and promising algorithm 
for the solution of
combinatorial optimization problems \cite{farhi2014quantum, Farhi2022quantumapproximate,Grange_2023,BLEKOS20241}, potentially more efficient than classical algorithms in terms of faster runtimes (time to find the optimal solution) or finding solutions of higher quality \cite{Shaydulin_2019, Sureshbabu2024parametersettingin,farhi25}. However, 
the 
standard variational QAOA ansatz, where the variational parameters of the quantum circuits are optimized in connection with a classical optimization routine, 
is not expected to lead 
to such a possible runtime advantage.
This can be understood from the
complexity of optimizing these parameters as follows:
the quantum circuit consists of $2p$ circuit parameters $(\vec{\gamma}, \vec{\beta})=(\gamma_1, ..., \gamma_p, \beta_1,...,\beta_p)$ that, in standard QAOA, all need to be optimized for a sufficiently large depth $p$. This requires a high number of circuit evaluations. Additionally, each evaluation itself contains a substantial number of shots (typically around $1000$ shots or more), to precisely determine the expectation value of the cost function, which is minimized by the variational algorithm. Furthermore, when using the variational ansatz, the classical optimizer can get stuck in local minima when optimizing $\vec{\gamma}$ and $\vec{\beta}$ \cite{kremenetski2021quantumalternatingoperatoransatz,PhysRevResearch.6.023171}, especially without a good initial guess. Therefore, fixed schedules, i.e., structured and predefined approaches for selecting circuit parameters, are of interest to circumvent expensive optimization routines \cite{Sureshbabu2024parametersettingin, PhysRevResearch.6.023171, brandao2018fixedcontrolparametersquantum}. It has been observed that the optimal parameters $(\vec{\gamma}^*, \vec{\beta}^*)$ exhibit patterns, namely that $\gamma_j^*$ increases smoothly while $\beta_j^*$ shows a decreasing behavior (with $j=1,...,p$) \cite{zhou2020}. These patterns are consistent with QAOA being interpreted as 
realizing a discrete adiabatic schedule with gradually changing
unitaries \cite{kremenetski2021quantumalternatingoperatoransatz}.

Based on these patterns, 
schedules can be formulated by which circuit parameters 
$(\vec{\gamma}, \vec{\beta})$
are determined for, in principle, arbitrarily high depth $p$.
Examples for such patterns are the Fourier heuristic strategy \cite{zhou2020}, Trotterized quantum annealing (TQA) \cite{Sack2021quantumannealing} or the linear-ramp schedule \cite{kremenetski2021quantumalternatingoperatoransatz, kremenetski2023quantumalternatingoperatoransatz, sakai2024linearlysimplifiedqaoaparameters, Shaydulin2021}, among others \cite{mbeng2019quantumannealingjourneydigitalization, crooks2018performancequantumapproximateoptimization}.
Sometimes (e.g., in \cite{zhou2020,Sack2021quantumannealing}), such schedules are used in order to provide initial parameters, which are then further optimized according to standard QAOA. Other works \cite{kremenetski2021quantumalternatingoperatoransatz, kremenetski2023quantumalternatingoperatoransatz, sakai2024linearlysimplifiedqaoaparameters, Shaydulin2021}
propose to use these schedules as fixed schedules, thus saving the enormous overhead involved in further optimizing a large number of circuit parameters, as explained above.

Indeed, a recent study claims that fixed linear-ramp schedules provide a runtime scaling advantage for some combinatorial optimization problems 
(in particular for a fully connected, randomly weighted MaxCut problem)
\cite{montanezbarrera2024universalqaoaprotocolevidence}. 
Apart from the QAOA depth $p$, the linear-ramp schedule exhibits only two free parameters $\Delta_\gamma$ and $\Delta_\beta$ (see Sec.~\ref{sec:LR-QAOA} below). The authors of Ref. \cite{montanezbarrera2024universalqaoaprotocolevidence} 
suggest that the parameter optimization needs to be performed only for one single problem instance of given size $N$, and then the same parameters can be used also for other instances of the same size.
Nevertheless, finding the optimal parameters even for a single problem instances at large $N$ constitutes a considerable effort.
The question remains how
suitable values of the parameters $\Delta_\gamma,\Delta_\beta$ and $p$ can be found in an efficient and scalable way. 
In our work, we address this question by proposing an extrapolation scheme whereby circuit parameters for larger problem sizes (in our case up to $N=28$) can be determined on the basis of optimizations performed only on subproblems of smaller size ($N=4,6,8,10$).
With these parameters, we present the scaling of the quantum runtime for four
combinatorial optimization problems: Portfolio Optimization, Feature Selection, Clustering 
and weighted MaxCut. The runtime 
is expressed
as the median of
the total depth of all quantum circuits that have to be executed to find the optimal solution.
After comparing the scaling of the quantum runtime with the classical runtime, we find indication of a quantum scaling advantage in the case of Portfolio Optimization.

Apart from the article \cite{montanezbarrera2024universalqaoaprotocolevidence} mentioned above, similar scaling advantages have recently been found also for Boolean satisfiability problems \cite{PRXQuantum.5.030348} and the low autocorrelation binary sequences (LABS) problem \cite{Shaydulin2024}. In both cases, fixed parameter schedules (though not as simple as linear-ramp) are considered, and the scaling of the QAOA runtime (combined with quantum minimum finding in \cite{Shaydulin2024}) is analyzed at fixed QAOA depth $p$.
In our work, we propose to adapt, both, the depth $p$ and also the circuit parameters $(\vec{\gamma}, \vec{\beta})$ with increasing problem size $N$. We therefore think it is possible that the QAOA runtime scalings obtained in the above mentioned works can be further improved when combining them with ideas presented in this article. 

This paper is divided into two main parts, a classical optimization and a quantum optimization part, and is organized as follows: in Sec. \ref{sec:COP}, we give a brief introduction to the combinatorial optimization problems considered in this paper, which are then addressed in Sec. \ref{sec:classical_optimization} by purely classical methods. This section provides the investigation as well as the discussion of the scaling behavior of different classical optimization methods (CPLEX, Gurobi, MQLib/Burer2002 Heuristic and Goemans-Williamson algorithm), which is evaluated by the time to find the optimal solution (runtime) as a function of the number of variables (bits). Sec. \ref{sec:qo} gives an introduction to the linear-ramp QAOA schedule and an explanation of our method to derive suitable QAOA parameters based on extrapolation from smaller towards larger number of qubits. The results are then discussed in Sec. \ref{sec:Results_and_Discussion} and we finally conclude in Sec. \ref{sec:Conclusion}.

\section{\label{sec:COP}Combinatorial Optimization Problems}
\noindent In this section, we introduce four
optimization problems that we all formulate as a QUBO  \cite {Glover2019} problem with the following structure:
\begin{equation}\label{eq:qubo}
    F(\mathbf{z}) = \sum_{i,j}^N F_{ij}z_i z_j + \sum_i^N f_i z_i
\end{equation}
with symmetric matrix $F$, vector $f$, number of bits $N$ and binary variables $z_1,...,z_N$.

Taking into account that $z_i z_i=z_i$ (for binary variables $z_i\in\{0,1\}$) the linear term can be included in the quadratic term, leading to
\begin{equation}\label{eq:qubo2}
    F(\mathbf{z}) = \sum_{i,j}^N Q_{ij}z_i z_j
\end{equation}
where  $Q_{ij}=F_{ij}+f_i\delta_{ij}$ is a symmetric matrix defining the problem under consideration.

Before discussing different combinatorial optimization problems, we want to stress that these examples are simplified compared to real-world problems in business and industry. While it would be possible to make them more realistic, it will be seen in the following that they nevertheless obtain generic features.

\subsection{\label{sec:Portfolioproblem}Formulation of the Portfolio Problem}

The portfolio optimization problem is one of the most famous and investigated topics in finance after Markowitz published his seminal paper \cite{Markowitz1952}. It addresses the question how to build a portfolio with different financial assets in order to maximize the return and to minimize the risk when the relation between these two properties is fixed. Whereas in banks and asset management firms a lot of subtleties have to been considered (e.g. use of sophisticated risk measures as Value-at-Risk, allowing for short-selling, use of integer asset fractions, transaction costs etc.), a generic formulation is obtained within the following framework: Use the historic return as a measure of the expected return, use the covariance of the asset returns as risk measure and restrict yourself to binary portfolio fractions. Using this together with the aforementioned QUBO function, see Eq.~\eqref{eq:qubo}, we can directly identify the cost function for the Portfolio optimization problem
\begin{equation}
    F_{\text{portfolio}}(\mathbf{z}) = q~\sum_{i,j}^N z_i z_j \sigma_{ij} - (1-q)~\sum_i^N z_i \mu_i
\end{equation}
with number of assets $N$, covariance matrix $\sigma_{ij}$, expected return vector $\mu_i$, binary portfolio weights $z_i=0$ or $1$ of stock $i$, and a risk preference factor $q$. 
A budget constraint $B=\sum_i z_i$ indicates the investment of a fixed budget, with $B$ being the number of assets chosen for the portfolio from $N$ total assets. A created portfolio is called feasible if the constraint is fulfilled; otherwise we call a portfolio infeasible. The constraint can be added to the cost function as a penalty term with a prefactor $A$:
\begin{equation}
    F_{\text{portfolio}}^{(A)}(\mathbf{z}) = F_{\text{portfolio}}(\mathbf{z}) + A\Bigl(\sum_{i=1}^N z_i - B \Bigr)^2 .\label{eq:FA}
\end{equation}
A suitable candidate for the prefactor $A$ can be computed using the routine described in Ref. \cite{Brandhofer2022}. In continuation of this paper, we use the annualized returns and covariance matrices for the German DAX 30 (as of 1/2/2021) between 01/01/2016 and 31/12/2020. 
For a given number of assets, different instances of the problem are obtained by randomly selecting 
$N$ out of the 30 DAX assets. The risk preference factor $q$ is typically set between $0$ and $1$ and, in contrast to our typical choice of $q=1/3$, it is set here to $q=1$, indicating the focus on the risk part of the problem irrespective of the return part.
Concerning the budget, we choose $B=N/2$ (for even $N$) or $B=(N-1)/2$ (for odd $N$).

\subsection{\label{sec:Feature Selection}Formulation of Feature Selection Problem}

Similarly, we can formulate the cost function for the feature selection problem in Machine Learning which can be based on two dependency measures: the first describes the dependence $\rho_{i,j}$ between the columns $i$ and $j$ of the feature matrix $X$, and the second is the dependence $\rho_{i,Y}$ between the feature column $i$ and the target vector $Y$. Furthermore, we consider a set of binary variables $z_i \in \{0,1\}$ for which $z_i=1$ indicates that the column $i$ is selected in the model, while $z_i=0$ refers to the column not being selected in the model. An intuitive approach is to select features such that the inter-feature dependence $\sum_{i,j=1}^N z_i z_j |\rho_{i,j}|$ is minimized and the dependence between feature and target $\sum_{i=1}^N z_i |\rho_{i,Y}|$ is maximized. The cost function can now be constructed as above by combining both criteria:\\
\begin{equation}
    F_{\text{feature}}(\mathbf{z}) = - \Bigl(\phi \sum_{i=1}^N z_i |\rho_{i, Y}| - (1-\phi)\sum_{\substack{i,j=1 \\ i \neq j}}^N z_i z_j |\rho_{i,j}| \Bigr)
\end{equation}
with weighting parameter $\phi \in [0,1]$, that we set to $\phi=0.9$. As a dependence measure between the features as well as between features and target we chose to take the linear correlation $\rho^{\text{Correl}}$ as proposed in \cite{HellsternDehnZaefferer2024}. In accordance to this reference, the “German Credit Data” \cite{germancreditdata} is used again. It comprises a total of 20 features, encompassing 7 numerical and 13 categorical attributes along with a an imbalanced binary target variable distinguishing non-defaulted cases (700 instances) from defaulted cases (300 instances).
After applying one-hot encoding to the categorical features, we end up with a maximum of 48 binary features.

\subsection{\label{sec:Clustering}Formulation of Clustering and randomly weighted MaxCut Problem}

In unsupervised learning, the problem of clustering data is defined as follows: Given a dataset {\bf D} containing $N$ data points, where each data point is represented as a feature vector $x_i$ in a feature space {\bf X}:
$D = \{x_1, x_2, ..., x_N\}$,
the goal of clustering is to partition the dataset into $K$ clusters, where $K$ is a user-specified or algorithmically determined parameter. Each cluster $C_k$ is a subset of the dataset {\bf D} with the following conditions:
\begin{enumerate}
    \item Completeness: Every data point $x_i$ belongs to exactly one cluster $C_k$, and the union of all clusters covers the entire dataset.

    \item  Homogeneity: Data points within the same cluster are similar to each other, and data points in different clusters are dissimilar. This is typically measured using a distance or similarity metric.

    \item Minimization Objective: The clusters are formed in a way that minimizes an objective function. This objective function may vary depending on the specific clustering algorithm being used. 
\end{enumerate}
Mathematically, the clustering problem can thus be defined as finding a partition of the dataset into K clusters that optimizes an objective function $J$, where: 
\begin{equation}
    J = f(D,C_1, C_2, ..., C_K) .
\end{equation}
The objective function $J$ can be chosen based on the specific clustering algorithm's criteria, such as minimizing intra-cluster distances or maximizing inter-cluster distances. 
The goal of clustering is to discover meaningful patterns or groupings within the data without any prior knowledge of the class labels. Various clustering algorithms, such as K-means \cite{lloyd1982least}, hierarchical clustering \cite{Sibson73}, and DBSCAN \cite{ester1996density}, can be used to solve this problem of finding the optimal partition of the dataset into clusters based on different criteria and distance measures.

The Euclidean distances for all data pairs $x_i,x_j$, i.e.
\begin{equation}
w_{ij}=||x_i-x_j||_2
\end{equation}
can be interpreted as the elements of a symmetric matrix $W$, where all diagonal elements are zero. 
This matrix $W$ can now be interpreted as an adjacency matrix of a graph $G$, where each vertex represents an element of $W$, and $w_{ij}$ is the weight of the edge between the vertices $i$ and $j$. In general, the graph is fully connected.

The common assumption in clustering -- distant points belonging to different clusters -- translates in the case of $K=2$ clusters to the problem of maximizing the overall sum of all weights (distances) between nodes with different labels. Therefore, we come to a Maximum-Cut
(MaxCut) problem - MaxCut(G, W) for the dense graph $G$ of the distance matrix $W$ \cite{Saiphet_2021, otterbach2017unsupervised}. A cut is a set of edges that separates all vertices $V$ into two disjoint sets $S$ and $\bar S$ with $S\cup \bar{S}=V$.
The cost of a cut is given by the sum of all weights of edges connecting vertices in $S$ with vertices in $\bar S$. With the introduction of binary variables $z_i \in \{0,1\}, i=1,...,N$  which denote whether $x_i$ belongs to $S$ or $\bar S$, the sum of all cuts can be expressed as the cost function
\begin{equation}
    F_{\text{clustering}}(\mathbf{z}) = \sum_{i,j=1}^N w_{ij}  (z_i+z_j-2\cdot z_i z_j) . 
\end{equation}

Again, we end up with a QUBO problem. Here, we decided to use the ``Moons'' and ``Blobs'' datasets of Scikit-Learn \cite{scikitlearn} with two features, i.e., $x\in \mathbb{R}^2$. Each data point then induces one binary decision variable. Using Moons and Blobs have the advantage that the clustering results can be either checked by e.g. the K-Means algorithm or visually. 

In addition to the MaxCut problem, that arises from the clustering problem formulation, we also consider MaxCut problems where the weights $w_{ij}$ of the edges are randomly drawn from a uniform discrete distribution between 0 and 1000 in steps of 1.

In Appendix \ref{boxplots_data} we show the distribution of non-diagonal and diagonal QUBO-matrix elements of the different optimization problems.

\section{Classical optimization}\label{sec:classical_optimization}

\begin{figure*}
\centering
\includegraphics[width=16cm]{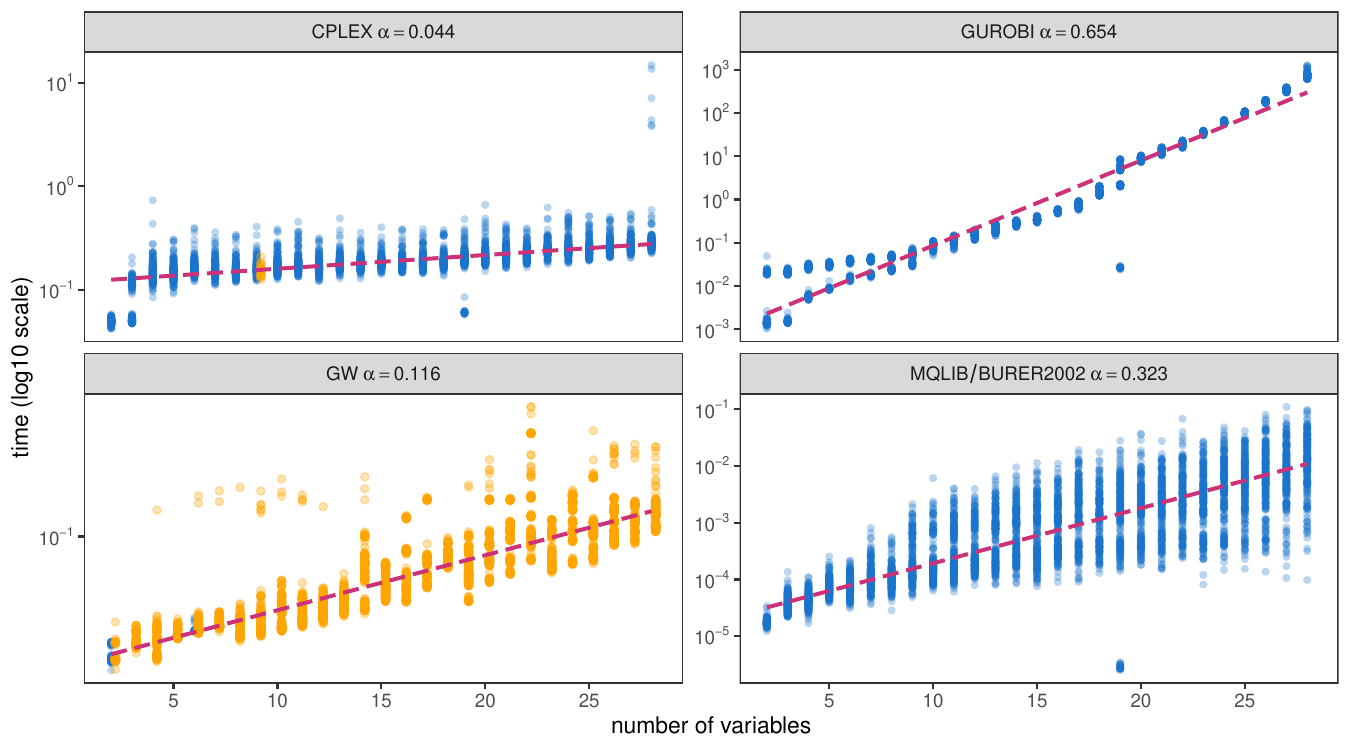}
\caption{Portfolio optimization: Runtime $T$ (seconds) obtained with CPLEX, Gurobi, Goemans-Williamson and MQLib/Burer2002, as a function of the number $N$ of variables (bits). The dots represent the results 
for the ten problem instances,
with blue indicating it was optimally solved, while orange indicates a non-optimal solution.
The slope $\alpha$ of a robust log2-linear model, i.e., $T\propto 2^{\alpha N}$ (red dashed line) is denoted in the header of each figure.}
\label{Portfolio_classical}
\end{figure*}

\begin{figure*}
\centering
\includegraphics[width=16cm]{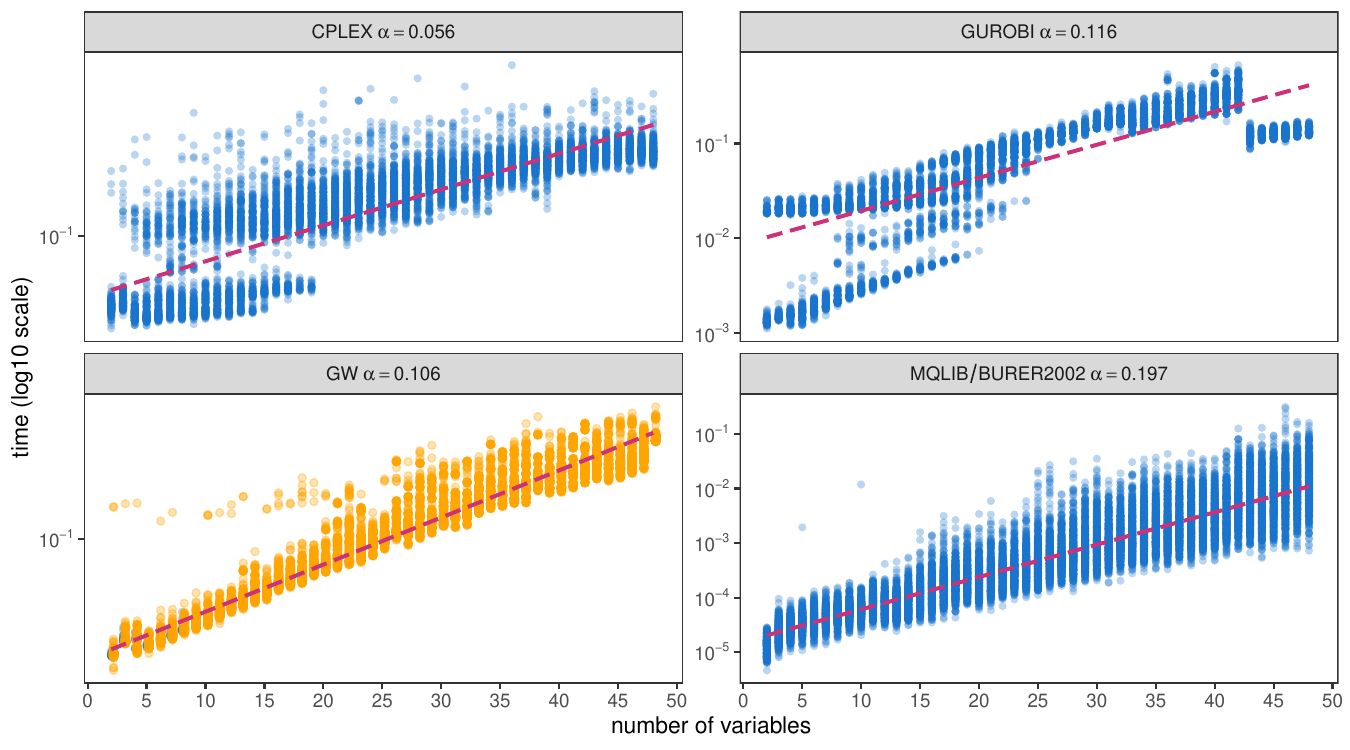}
\caption{Feature Selection: Runtime $T$ (seconds) obtained with CPLEX, Gurobi, Goemans-Williamson and MQLib/Burer2002, as a function of the number $N$ of variables (bits). The dots represent the results 
for the ten problem instances,
with blue indicating it was optimally solved, while orange indicates a non-optimal solution.
The slope $\alpha$ of a robust log2-linear model, i.e., $T\propto 2^{\alpha N}$ (red dashed line) is denoted in the header of each figure.}
\label{Feature_classical}
\end{figure*}

\begin{figure*}
\centering
\includegraphics[width=16cm, height=8.5cm]
{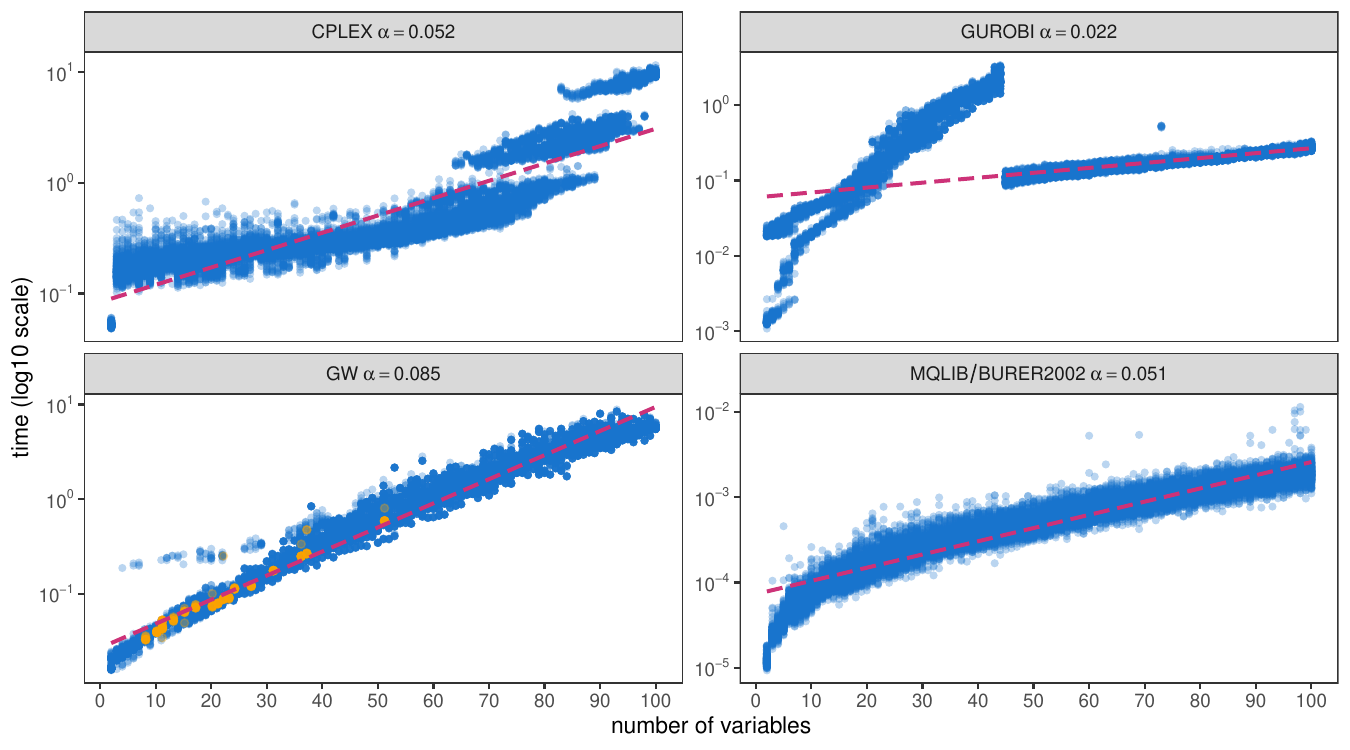}
\caption{Clustering (Moons): Runtime $T$ (seconds) obtained with CPLEX, Gurobi, Goemans-Williamson and MQLib/Burer2002, as a function of the number $N$ of variables (bits). The dots represent the results 
for the ten problem instances,
with blue indicating it was optimally solved, while orange indicates a non-optimal solution.
The slope $\alpha$ of a robust log2-linear model, i.e., $T\propto 2^{\alpha N}$ (red dashed line) is denoted in the header of each figure.}
\label{Moons_classical}
\end{figure*}

\begin{figure*}
\centering
\includegraphics[width=16cm, height=8.5cm]
{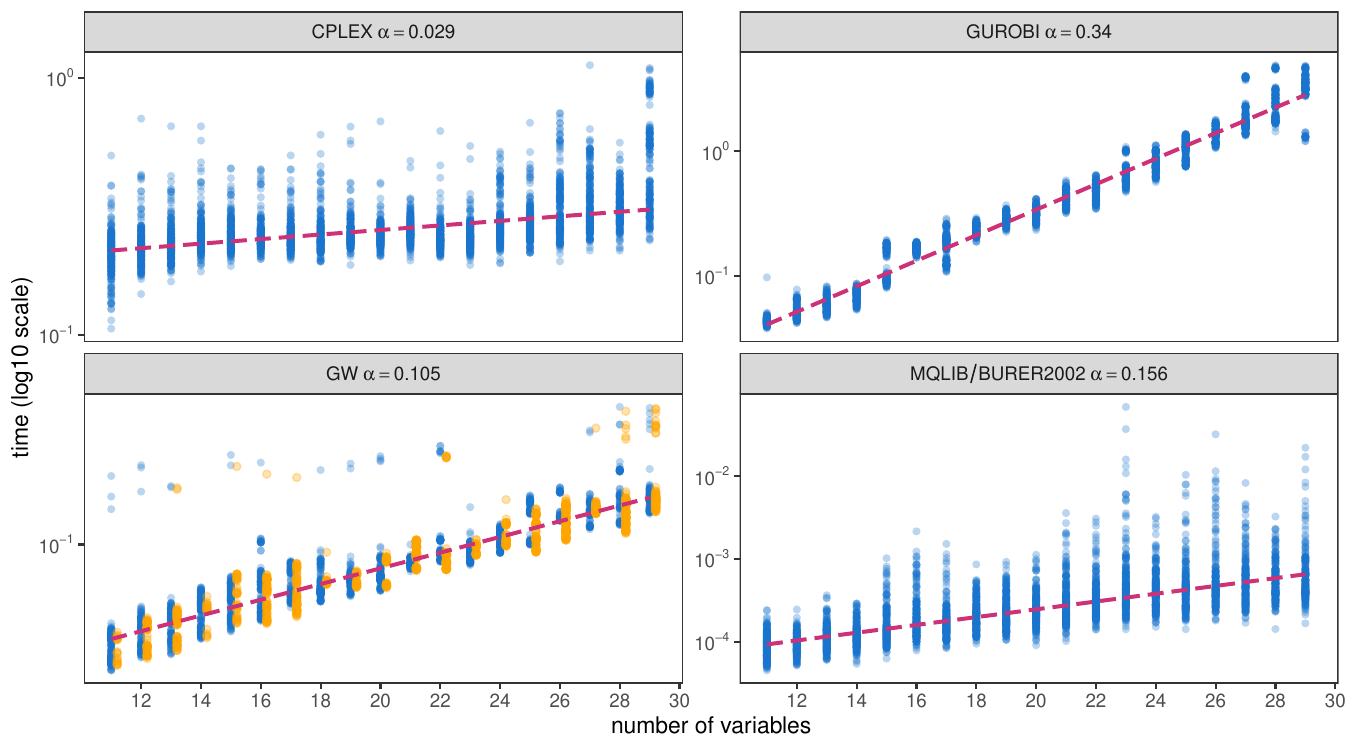}
\caption{MaxCut: Runtime $T$ (seconds) obtained with CPLEX, Gurobi, Goemans-Williamson and MQLib/Burer2002, as a function of the number $N$ of variables (bits). The dots represent the results of individual algorithm runs on one instance, with blue indicating it was optimally solved, while orange indicates a non-optimal solution. The slope $\alpha$ of a robust log2-linear model, i.e., $T\propto 2^{\alpha N}$ (red dashed line) is denoted in the header of each figure.}
\label{maxcut_classical}
\end{figure*}

To solve the optimization problems introduced above, we use the following methods/software solutions:
\begin{itemize}
\item CPLEX \cite{cplex2009v12}
\item Gurobi \cite{gurobi}
\item MQLib/Burer2002 Heuristic \cite{DunningEtAl2018}
\item Goemans-Williamson's algorithm \cite{Goemans1995}
\end{itemize}
Apart from MQLib/Burer2002, these classical optimization methods are all available in IBM's Qiskit \cite{Qiskit} for solving QUBOs with a uniform function call. Due to the large number of all possible solutions ($2^N$), a brute-force approach is infeasible for large $N$. 
Note that Goemans-Williamson's is an approximative algorithm, 
it does not guarantee to find the global optimum.
In the literature, also more physics-inspired optimization methods are described, such as the parallel-tempering with isoenergetic cluster moves approach, see \cite{PhysRevE.99.063314}. We refrained from implementing these approaches because our aim was primarily to use industry established and ready-to-use optimization frameworks for our work.

Strictly speaking, none of the algorithms guarantees to find the global optimum, although at least CPLEX and Gurobi may provide additional information about the optimality of solutions. That is, they compute upper and lower bounds for the objective function value, which may indicate global optimality if the difference between both bounds is zero. 
MQLib/Burer2002, finally, does not provide information about the optimality of the solution. It continues trying to find better solutions until a user-defined maximum runtime is reached. At the end, it returns these solutions together with the times at which they have been found.

To investigate the scaling behavior of the different classical optimization methods, we perform the following experiments:

\begin{enumerate}
    \item Sample $I$ instances with size $N$ randomly from the respective data set/distribution, for each combinatorial optimization problem class. This accounts for variation of the runtime caused by instance characteristics.
    \item For each instance, we measure $R$ times how long each algorithm takes to find a solution. The repetition is intended to account for random variations of the runtime.
    \item Repeat this for each $N\in \{N_{\text{min}}, N_{\text{min}}+1, \dots, N_{\text{max}}\}$. This accounts for the influence of the problem size.
\end{enumerate}
The specific settings considered for each problem class are:
\begin{itemize}
    \item Portfolio Optimization: $I=10$, $R=20$, $N \in \{2,3,\dots,28\}$
    \item Feature Selection: $I=20$, $R=10$, $N \in \{2,3,\dots,48\}$
    \item Clustering (Moons): $I=20$, $R=10$, $N \in \{2,3,\dots,100\}$
    \item Clustering (Blobs): $I=20$, $R=10$, $N \in \{2,3,\dots,100\}$
    \item MaxCut: $I=10$, $R=20$, $N \in \{11,12,\dots,29\}$
\end{itemize}
 In the following, we show the runtime measured during execution as function of the number $N$ of variables and the slope of the fit through the data. The slope is an estimator of the scaling factor which depends on the data set and on the algorithm.

In Fig. \ref{Portfolio_classical} the results for the Portfolio Optimization problem are shown. The absolute values for the runtime differ strongly between the four classical optimization schemes. On the one hand, Gurobi shows the longest  absolute runtimes as well as the highest scaling factor. On the other hand, CPLEX shows the lowest scaling factor and MQLib/Burer2002 the shortest absolute runtime. 

A similar behavior can be observed for the Feature Selection problem, see Fig. \ref{Feature_classical}. Again, CPLEX has the lowest scaling factor, MQLib/Burer2002 the shortest absolute runtimes, and Goemans-Williamson's algorithm does not find the optimal solution in most cases. However, MQLib/Burer2002 now has the highest scaling factor, and GW/CPLEX the longest absolute runtimes. 

In Fig. \ref{Moons_classical} the Clustering results for the Moons dataset are shown. 
Interestingly, GW mostly finds the optimal solutions for this problem class, and Gurobi now has the lowest scaling factor, albeit also showing a strong discontinuity of runtimes at about $N=45$. 
Similarly, but less pronounced, 
CPLEX also
displays jumps of the runtime, e.g., around $N=90$.
GW has the largest scaling factor in this case. MQLib/Burer2002 has a marginally lower scaling factor than CPLEX, and is again the algorithm with lowest absolute runtimes.

The results for the Blobs dataset are depicted in Fig. \ref{Blobs_classical} which can be found in the Appendix \ref{class_data}.
In this case, CPLEX has the lowest $\alpha$, closely followed by Gurobi and MQLib/Burer2002 (with lowest absolute runtimes). Besides some instances with very small $N$, CPLEX does not show the discontinuities displayed on the Moons data while Gurobi has the same jump at about $N=45$, in addition to jumps at lower values of $N$.

Compared to the other problem classes, the MaxCut instances seem to exhibit the least number of discontinuities, as shown in Fig.~\ref{maxcut_classical}. GW only finds the optimal solution in some cases. In terms of our scaling estimates $\alpha$ of the other three algorithms, CPLEX performs best, followed by MQLib/Burer2002, while Gurobi has the largest scaling factor. In terms of absolute values, MQLib/Burer2002 has again the lowest absolute runtimes.

For each problem instance, we also performed separate runs with CPLEX (focusing on optimality), to confirm whether we find globally optimal reference solutions in each case. CPLEX provides an upper bound of the gap to the global optimum, which can become zero, in which case the corresponding reference solution that was found is provably a global optimum. Overall, this gives us only a small number of instances for which we have no proof that the best solution we found is globally optimal:
\begin{itemize}
    \item Portfolio Optimization: 1 of 270 instances
    \item Feature Selection:  18 of 940 instances
    \item Clustering (Moons): 2 of 1980 instances
    \item Clustering (Blobs): 4 of 1980 instances
    \item MaxCut:  0 of 190 instances
\end{itemize}
Moreover, in all cases, the relative gap between the
best-found solution and the lower bound of the objective function value is smaller than $10^{-6}$.
Note, that in the figures \ref{Portfolio_classical} - \ref{Moons_classical}, blue dots mark a run as solved even if we have no proof that the solution is a global optimum. Orange dots imply that the algorithm found a solution that is worse than the reference we determined.

In summary, it can be concluded that the various classical optimization methods exhibit different scaling behaviors on the investigated combinatorial optimization problems. For the subsequent comparison with quantum algorithms, we will utilize the MQLib/Burer2002 algorithm for the following reasons: The Goemans-Williamson's algorithm is unsuitable for the comparison because it frequently does not find the optimum. The commercial optimization algorithms CPLEX and Gurobi are problematic because their search approach may change unexpectedly depending on various criteria, as is also evidenced by the jumps and discontinuities of their runtime graphs, making it very hard to model their behavior. 
Moreover, the runtimes of CPLEX and Gurobi also include the efforts required for collecting additional information such as optimality bounds, which neither MQLib/Burer2002 nor the quantum algorithm discussed below do provide. Especially for small problem sizes, the runtimes of CPLEX (of the order of $10^{-1}$s) exhibit a considerable overhead compared to MQLib/Burer2002 (about $10^{-5}$ or $10^{-4}$s).
For these reason, we argue that only the MQLib/Burer2002 algorithm reliably reflects the actual algorithmic complexity for finding the optimal solution and therefore appears to be the most suitable for comparison with the quantum algorithm.

\section{\label{sec:qo}Quantum optimization}
\subsection{\label{sec:LR-QAOA}Linear-ramp QAOA}

\begin{figure}[!h]
\centering
\includegraphics[width=8cm]
{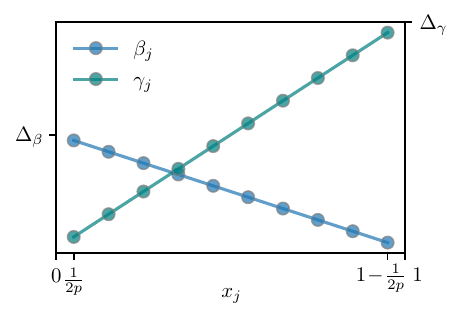}
\caption{Linear-Ramp protocol according to Eqs.~(\ref{eq:linear_ansatz}-\ref{eq:linear_ansatz_xj}) \\(illustrated here for $p=10$).}
\label{fig:lr_protocol}
\end{figure}

When using QAOA to solve the above introduced optimization problems on a quantum computer, the following parameterized quantum state is generated: 
\begin{equation}
    \ket{\psi_{\vec{\gamma}, \vec{\beta}}} = \hat{U}_M(\beta_p) \hat{U}_F(\gamma_p) ... \hat{U}_M(\beta_1) \hat{U}_F(\gamma_1) \ket{\psi_0}_M
\label{eq:QAOA_circuit}
\end{equation}
with circuit parameters $\vec{\gamma}=(\gamma_1,...,\gamma_p)$ and $\vec{\beta}=(\beta_1,...,\beta_p)$, mixer unitary $\hat{U}_M(\beta_j)=e^{-i\beta_j\hat{M}}$ with transverse-field mixer $\hat{M}=-\sum_{i=1}^n \hat{X}_i$, and cost unitary $\hat{U}_F(\gamma_j)=e^{-i\gamma_j\hat{F}}$ with cost Hamiltonian $\hat{F}$ (defined by $\hat{F}|\mathbf{z}\rangle = 
F(\mathbf{z}) |\mathbf{z}\rangle$ for basis states $|\mathbf{z}\rangle$). The initial state $\ket{\psi_0}_M$ corresponds to the mixer Hamiltonian $\hat{M}$, or more precisely, it is its eigenstate associated with its smallest eigenvalue. Hence, the state is the equal superposition state $\ket{\psi_0}=\ket{+}^{\otimes N}$ for this choice of $\hat{M}$.
Finally, the probability of measuring the optimal solution $\mathbf{z_{\rm opt}}$ is obtained as:
\begin{equation}
P_\text{opt} = \left|\langle \mathbf{z_{\rm opt}}|\psi_{\vec{\gamma}, \vec{\beta}}\rangle\right|^2
\label{eq:Popt}
\end{equation}
In contrast to standard QAOA, where suitable values for the circuit parameters are suggested through a classical optimization routine, we sample them directly from the ansatz 
\begin{eqnarray}
    \gamma_j & = &  \Delta_\gamma x_j\label{eq:linear_ansatz}\\
    \beta_j & = & \Delta_\beta (1-x_j)
\end{eqnarray}
with
\begin{equation}
    x_j = \frac{j-\frac{1}{2}}{p},\ \ j=1,2,\dots,p,\label{eq:linear_ansatz_xj}
\end{equation}
where the parameters $\Delta_{\gamma}, \Delta_{\beta}$ are left to be optimized. Additionally, we find optimized values for $p$, with the procedure outlined in more detail in Sec. \ref{sec:evaluation}. Those three parameters characterize the linear-ramp schedule which is illustrated in Fig. \ref{fig:lr_protocol}. 

Intuitively, the above linear-ramp schedule can be motivated by the quantum adiabatic theorem \cite{born28}, where the known ground state of $\hat{M}$ is transformed to the desired ground state of $\hat{F}$. Indeed, the above QAOA circuit, Eq.~(\ref{eq:QAOA_circuit}) reproduces the continuous adiabatic evolution in the limit $\Delta_\gamma,\Delta_\beta\to 0$ and  $\Delta_\gamma p,\Delta_\beta p\to \infty$.
As it turns out, however, the continuous limit is not necessary in order to prove convergence to the ground state (i.e. $P_{\rm opt}\to 1$) in the limit $p\to\infty$: according to the discrete adiabatic theorem \cite{dranov98}, this convergence is also guaranteed for finite $\Delta_\gamma,\Delta_\beta>0$
-- provided that $\Delta_\gamma,\Delta_\beta$ do not exceed a critical value $\Delta_{\rm crit}$ in order to prevent the eigenvalues of the unitary QAOA operators to "wrap around" the unit circle in the complex plane \cite{kremenetski2023quantumalternatingoperatoransatz}. (In our case, $\Delta_{\gamma,{\rm crit}}=2\pi/(F_{\rm max}-F_{\rm min})$ and $\Delta_{\beta,{\rm crit}}=\pi/N$, with $F_{\rm min}$ and $F_{\rm max}$ the smallest and largest eigenvalues of $\hat{F}$.)

From a practical point of view, however, we are rather interested in the behaviour of LR-QAOA at finite $p$ instead of $p\to\infty$. Even if the limit $P_{\rm opt}\to 1$ is not reached in this case, the optimal solution can still be found by repeating the circuit several (approximately $1/P_{\rm opt}$) times. In the following, we will quantify the time required to find the optimal solution on a quantum computer in terms of the total depth $D$, defined as the depth of a single QAOA circuit (which, in turn, is proportional to $p$) times the number of circuit repetitions, see Eqs.~(\ref{eq:D},\ref{eq:Nshots}) below. 

\subsection{\label{sec:Extrapolation}LR-Extrapolation}

We follow the idea to discard the variational optimization of the circuit parameters and finding suitable values for $\Delta_{\gamma}, \Delta_{\beta}$ and $p$ by extrapolating from smaller to larger problem sizes, corresponding to the number of qubits $N$. 
This enables us to make a prediction of the circuit parameters for larger problem sizes, while only the optimization over small problem sizes has to be performed. The method consists of three parts: the first part includes a reduction of the original problem to smaller subproblems, the second part describes the extrapolation,
and the third part focuses on the post-processing of the results. 

\subsubsection{\label{sec:Random Sampling}Random Sampling and Grid Search}

For each QUBO instance of size $N\in \{12,...,28\}$, we first create random sub-instances of size $\tilde{N}\in \{4,6,8,10\}$. This is done by randomly selecting $\tilde{N}$ out of the $N$ binary variables and then constructing the QUBO cost functions as described in  
Sec.~\ref{sec:COP}. For Feature Selection and Clustering, this simply amounts to choosing the corresponding sub-matrices of $Q_{ij}$, see Eq.~\eqref{eq:qubo2}. For Portfolio Optimization, the dependence of the budget constraint $B$ on the problem size is taken into account (whereas the penalty term $A$ is kept constant), see Eq.~\eqref{eq:FA}. Thereby, 10 random sub-instances are created for each $\tilde{N}$. 

In the next step, we use the 10 sampled sub-instances of the smallest size ($\tilde{N}=4$) and perform a grid search with an initial depth value $p_0 = 6$.
That is, we construct a grid of $\log_{10}(\Delta_{\gamma})$, $\log_{10}(\Delta_{\beta})$ points, compute the probability of finding the optimal solution $P_{\text{opt}}$ for each instance and determine the average probability $\overline{P_{\text{opt}}}$ (averaged over the 10 sub-instances). For all subproblem sizes, we choose a $11\times 11$ grid with 
$\log_{10}(\Delta_\gamma)_\text{min} \leq \log_{10}(\Delta_\gamma) \leq \log_{10}(\Delta_\gamma)_\text{max}$ and $\log_{10}(\Delta_\beta)_\text{min} \leq \log_{10}(\Delta_\beta) \leq \log_{10}(\Delta_\beta)_\text{max}$.
The initial size of the 
grid (for $\tilde{N}=4$) is chosen as 
$\log_{10}(\Delta_\gamma)_\text{max}-\log_{10}(\Delta_\gamma)_\text{min}= \log_{10}(\Delta_\beta)_\text{max}-\log_{10}(\Delta_\beta)_\text{min}=2$, centered around
$[\log_{10}(\Delta_\beta)_\text{min}+\log_{10}(\Delta_\beta)_\text{max}]/2=0.5$ in case of $\log_{10}(\Delta_\beta)$,
whereas the center of the $\log_{10}(\Delta_\gamma)$-grid depends on the particular class of problem instance (due to different scalings of the corresponding QUBO matrices, see Appendix~\ref{boxplots_data}). In particular, we choose for
$[\log_{10}(\Delta_\gamma)_\text{min}+\log_{10}(\Delta_\gamma)_\text{max}]/2$ the values $0$ for portfolio optimization and clustering, $+2$ for feature selection, and $-2.5$ for MaxCut.
The logarithmic scale, i.e., considering $\log_{10}(\Delta_{\gamma,\beta})$ instead of $\Delta_{\gamma,\beta}$, is used to cover a sufficiently wide range of values. 

\subsubsection{Multivariate Skew Gaussian Fit and Grid Reduction}

\begin{figure}
\centering
\includegraphics[width=8cm]
{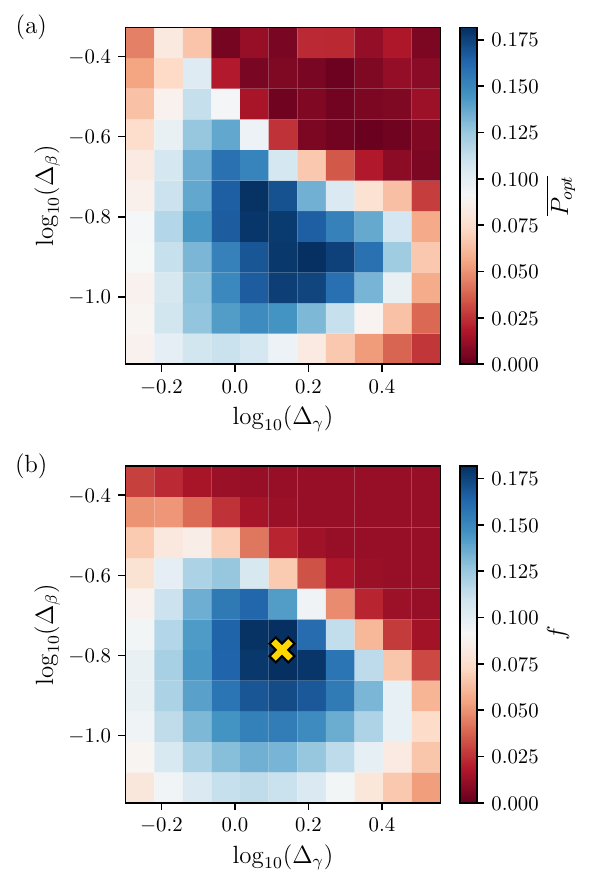}
\caption{(a) Portfolio Optimization: 2-dimensional grid search landscape for 10 sub-problems
of size $\tilde{N}=10$ sampled from a problem of size $N=28$ at depth $p=107$. The blue region shows the region with highest average ground state probability $\overline{P_\text{opt}}$. (b) Skewed Gaussian fit $f$, see Eq.~(\ref{eq:fit}), to the data shown in (a). The optimal values ($\log_{10}(\Delta_{\gamma,\text{opt}})$, $\log_{10}(\Delta_{\beta,\text{opt}})$) are represented by the yellow cross. 
}
\label{fig:gridsearch_skewed_fit}
\end{figure}

The above grid search provides a two-dimensional landscape consisting of $11\times 11=121$ data points, see Fig.~(\ref{fig:gridsearch_skewed_fit}) as an example (for sub-problem size $\tilde{N}=10$ and depth $p=107$). We can clearly see a dark blue region where the average probability  $\overline{P_{\text{opt}}}$ assumes the largest values. It is the center of this region that we are interested in and which we will use for our extrapolation procedure below.

In order to determine the position of the center, we fit the discrete $11\times 11$-landscape by a continuous function. 
Empirically, we  found that the following  multivariate skew Gaussian function is well suited in order to approximate the $\overline{P_{\text{opt}}}$-landscape (the detailed structure of which in principle depends on the random choice of sub-instances) by a simple and smooth function:
\begin{align} \label{eq:fit}
    f_{\bm{\mu},\Sigma,\bm{\alpha},A,B}(\textbf{x}) = A~\text{exp}\Bigl(-\frac{1}{2} (\textbf{x}-\bm{\mu})^T \Sigma^{-1} (\textbf{x}-\bm{\mu})\Bigr) \\ \times \frac{1}{2}\left[1 + \text{erf}\left(\bm{\alpha}\cdot\frac{\textbf{x}-\bm{\mu}}{\sqrt{2}}\right)\right]+B\nonumber
\end{align}
where $\textbf{x}=\{\log_{10}(\Delta_\gamma),\log_{10}(\Delta_\beta)\}$. The fit parameters consist of a two-dimensional vector $\bm{\mu}$ (center of the Gaussian function), a $2\times 2$ (inverse) covariance matrix $\Sigma^{-1}\geq 0$, a two-dimensional vector $\bm{\alpha}$ characterizing the skewness and two parameters
$A,B\geq 0$. The maximum of this function is determined numerically and provides us with $\{\log_{10}(\Delta_{\gamma,\text{opt}}),\log_{10}(\Delta_{\beta,\text{opt}})\}=
\textbf{x}_\text{opt}$ and $\left.\overline{P_\text{opt}}\right|_\text{max}=f_{\bm{\mu},\Sigma,\bm{\alpha},A,B}(\textbf{x}_\text{opt})$.
Additionally, we extract the widths $s_\gamma = \sqrt{\Sigma_{1,1}}$ and
$s_\beta = \sqrt{\Sigma_{2,2}}$ of the blue region from the diagonal elements of the covariance matrix $\Sigma$. 

With these parameters, we set the new coordinates for the grid search of the next-sized subproblem in the following way
\begin{align}
    \log_{10}(\Delta_\gamma)_\text{min} = \log_{10}(\Delta_{\gamma,\text{opt}}) - s_\gamma \\ \nonumber
    \log_{10}(\Delta_\gamma)_\text{max} = \log_{10}(\Delta_{\gamma,\text{opt}}) + s_\gamma \\ \nonumber
    \log_{10}(\Delta_\beta)_\text{min} = \log_{10}(\Delta_{\beta,\text{opt}}) - s_\beta \\ \nonumber
    \log_{10}(\Delta_\beta)_\text{min} = \log_{10}(\Delta_{\beta,\text{opt}}) + s_\beta \\ \nonumber
\end{align}
which ensures that the optimum of the previous-sized problem is at the center of the following grid. This procedure is then repeated until we reach sub-problem size $\tilde{N}=10$. 

\begin{figure*}
\centering
\includegraphics[width=0.95\textwidth]
{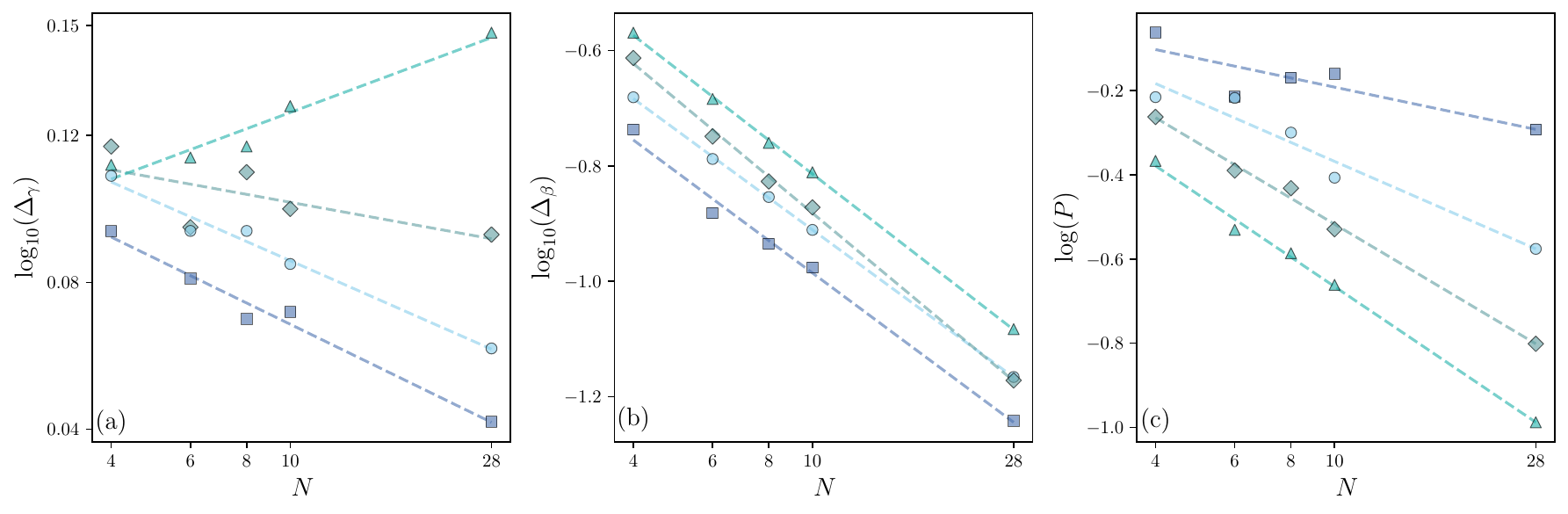}
\caption[width=1.0\textwidth]{Extrapolation of (a) log($\Delta_{\gamma}$), (b) log($\Delta_{\beta}$) and (c) log($P_\text{opt}$)
for four instances of the portfolio optimization problem with $N=28$ assets
(triangles, diamonds, circles, squares). 
The symbols at $N=4,6,8,10$ indicate the optimized values 
found for the respective subproblems. The dashed lines represent linear fits (on log-log scale) through these points. Finally, the markers at 
$N=28$ display the extrapolated values of these parameters.   Each instance is processed at a certain depth: $p=128$ (triangles), $p=215$ (diamonds), $p=304$ (circles), $p=608$ (squares).}
\label{fig:m2_extrapolation}
\end{figure*}

\subsubsection{\label{subsec:extrapolation}Extrapolation}

After having found optimized parameters
$\{\log_{10}(\Delta_{\gamma,\text{opt}}),\log_{10}(\Delta_{\beta,\text{opt}})\}$ and $\left.\overline{P_\text{opt}}\right|_\text{max}$ for all sizes
$\tilde{N}=4,6,8,10$ of the sub-instances,
we extrapolate these values towards the size $N$ of the original instance.
For this purpose, we assume a linear behavior on a log-log scale,
i.e., we fit functions $y_i = a_i x + b_i$ 
to the above data points, where $x=\log_{10}(\tilde{N})$, $y_1=\log_{10}(\Delta_{\gamma,\text{opt}})\}$,
$y_2 = \log_{10}(\Delta_{\beta,\text{opt}})$ and 
$y_3 = \log_{10}\left(\left.\overline{P_\text{opt}}\right|_\text{max}\right)$.
Thereby, we obtain extrapolated values
$\Delta_{\gamma,\text{extr}}$, $\Delta_{\beta,\text{extr}}$ of the linear-ramp circuit parameters
together with an estimation $P_\text{opt,extr}$ of the probability to measure the optimal solution
of our problem instance with size $N$.
In Fig. \ref{fig:m2_extrapolation}, this extrapolation procedure is illustrated for four different portfolio optimization instances of size $N=28$, each at a certain depth $p$ ($128$ ,$215$, $304$ and $608$). Each depth corresponds to the optimal depth for that instance, as determined in Sec. \ref{sec:evaluation} below.

\begin{figure}
\centering
\includegraphics[width=9cm]
{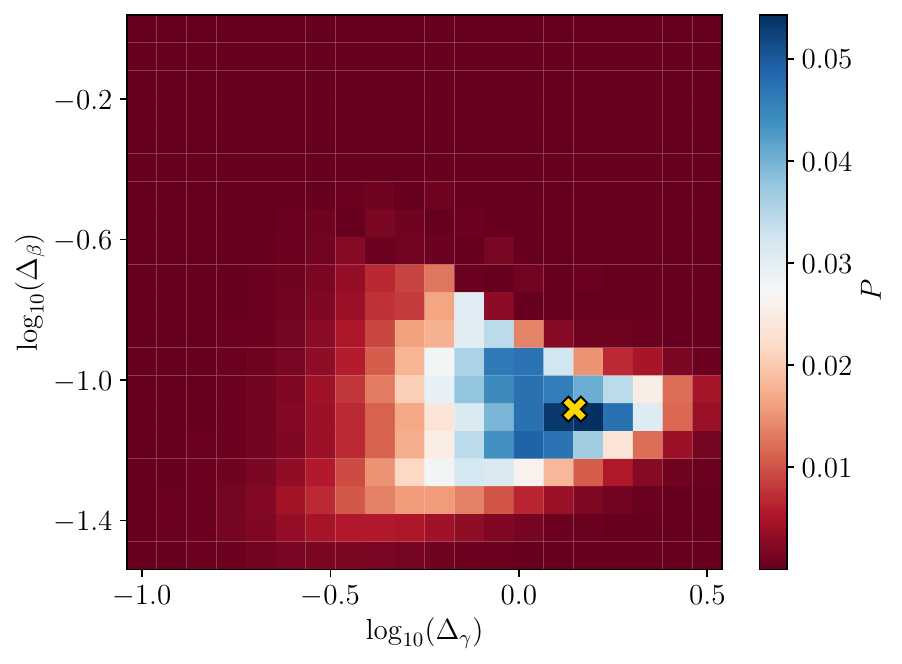}
\caption{Grid search landscape for one instance (corresponding to triangles in Fig.~\ref{fig:m2_extrapolation}) with $N=28$ qubits and QAOA depth $p=128$. 
The optimal region is visualized by the dark blue region. The optimal $(\Delta_{\gamma,\text{extr}}, \Delta_{\beta,\text{extr}})$ configuration predicted by the extrapolation method, represented by the yellow `x' symbol, is close to the optimal configuration found through grid search. 
Moreover, the corresponding maximum value of $P_\text{opt}\simeq 0.055$ (see the labels of the color bar on the right-hand side) 
is of the same order of magnitude as
the extrapolated value of $P_\text{opt,extr}\simeq 0.103$.
This indicates that the extrapolation method performs well.}
\label{fig:gridsearch_N28_w_extopt}
\end{figure}

At present, we do not have a theoretical justification for the algebraic scaling with respect to $\tilde{N}$ assumed above (i.e. linear behaviour on log-log scale), but we empirically found that it provides better results than exponential scaling, for example.
To check the quality of our extrapolation, we have determined
a $20\times 20$ grid search landscape for one of the four instances shown in Fig.~\ref{fig:m2_extrapolation}. Note that this grid search is numerically quite expensive,
since it involves simulations of large quantum circuits (with $N=28$ qubits) -- in contrast to the extrapolation procedure based on simulations of small circuits (with $\tilde{N}=4,6,8,10$).
As shown in Fig.~\ref{fig:gridsearch_N28_w_extopt}, the estimated values
$\Delta_{\gamma,\text{extr}}$, $\Delta_{\beta,\text{extr}}$ obtained from the extrapolation agree quite well with the true maximum. This example suggests that the predictions made using the extrapolation method provide a good estimate for the linear-ramp parameters.
More data supporting this conclusion (at least for most instances) will be shown in Sec.~\ref{sec:Results_and_Discussion}.

\subsubsection{Total depth and optimization of $p$}\label{sec:evaluation}

\noindent As described above, the extrapolation procedure provides us with suitable estimates of the
linear-ramp circuit parameters $\Delta_{\beta,\gamma}$ and the corresponding probability
$P_\text{opt}$ of measuring the optimal solution. What remains is to choose an appropriate value of the QAOA depth $p$. In principle, we may increase  $P_\text{opt}$ by increasing $p$, which, however, leads to longer circuits. To find an optimal value of $p$, we want to minimize the 
runtime of our quantum algorithm until the optimal solution is found. 
For this purpose, we concentrate in the following on the depth $d$ of the QAOA quantum circuit.
The corresponding time is then given by multiplying the depth with the time of a single quantum gate. Depending on the given quantum hardware, more specific estimates could be obtained, e.g., by taking into account different times of single-qubit and two-qubit gates or times of initial state preparations and final state measurements. We expect, however, that these details will not significantly affect the scaling behavior in the limit of large problem sizes.

To determine the total quantum runtime, we therefore multiply the depth $d$ of a single QAOA circuit (which, in turn is roughly proportional to the QAOA depth $p$) with the number $N_\text{shots}$ of shots required to find the optimal solution:
\begin{equation}
    D = d  N_{\text{shots}}
    \label{eq:D}
\end{equation}
The median of the latter (defined by the condition that the probability to find the optimal solution after $N_\text{shots}$ circuit executions equals $1/2$) is given by:
\begin{equation}
    N_\text{shots}=\frac{\text{log}(\frac{1}{2})}{\text{log}(1-P_{\text{opt}})}
    \label{eq:Nshots}
\end{equation}
Alternatively, the expectation value given by $1/P_\text{shots}$ could also be used instead of the median.

\begin{figure*}
\centering
\includegraphics[width=16cm]
{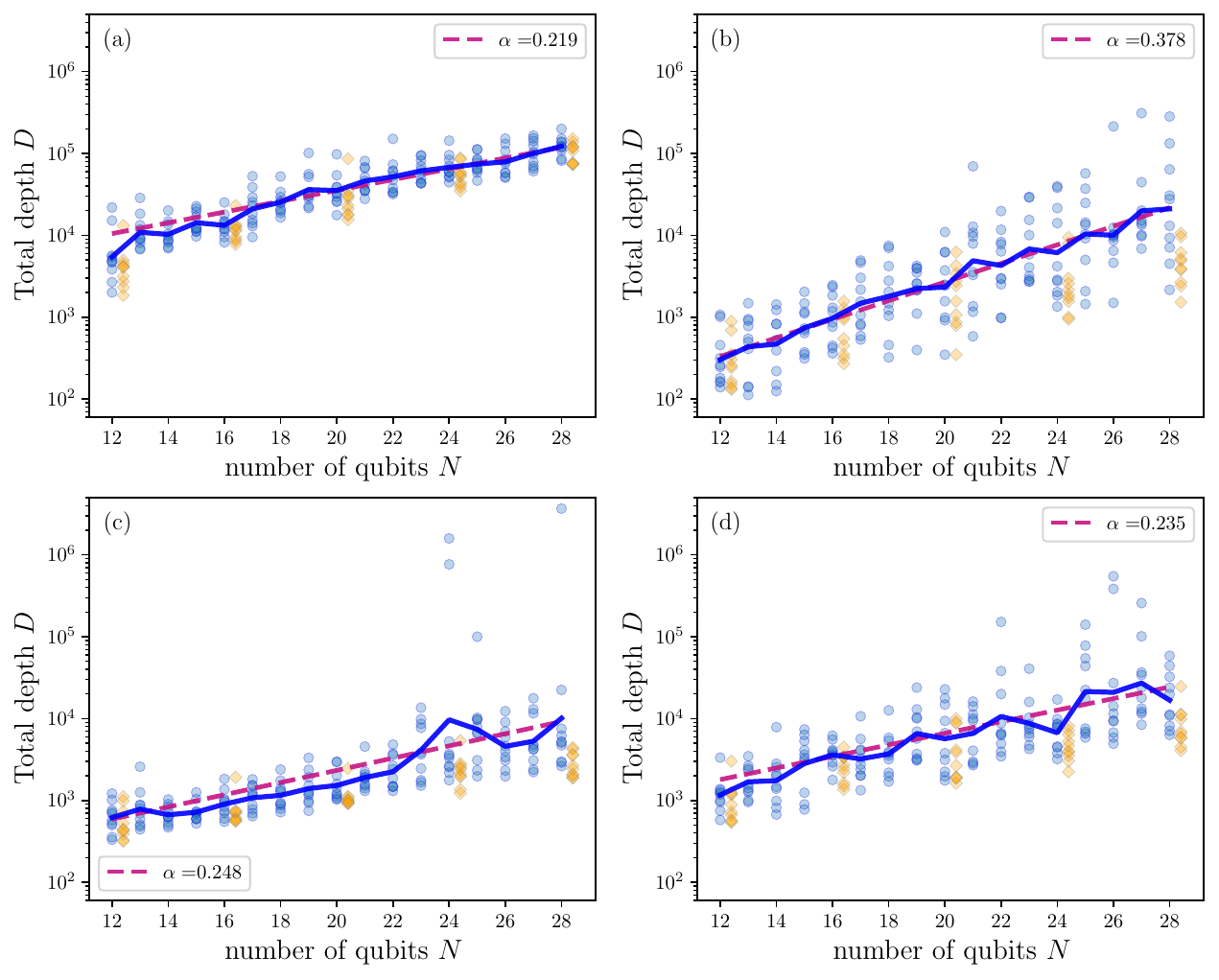}
\caption{Scaling of quantum runtime expressed as total depth $D$, see Eq.~(\ref{eq:D}), with increasing problem size $N$ (number of qubits) using linear-ramp QAOA with circuit parameters determined from extrapolation individually for each instance. The red dashed lines represent exponential fits with slope $\alpha$ (i.e. $D \propto 2^{\alpha N}$) of the geometric means (blue lines) over 10 problem instances per $N$ (transparent blue dots). (a) Portfolio Optimization. Compared to the classical runtime $\alpha_{\rm cl}=0.323$, see Fig.~\ref{Portfolio_classical} (MQLib/Burer2002),
the increase of the quantum runtime is slower ($\alpha=0.219$), thus indicating a potential quantum scaling advantage. For the remaining three cases (b) Feature Selection ($\alpha=0.378$ vs. $\alpha_{\rm cl}=0.197$), (c) Clustering (Moons) ($\alpha=0.248$ vs. $\alpha_{\rm cl}=0.051$), and (d) MaxCut ($\alpha=0.235$ vs. $\alpha_{\rm cl}=0.156$), a scaling advantage compared to the classical runtime is not observed. The yellow diamond shapes represent the optimal values found by a minimizer, starting from the extrapolated parameters as initial point. For better visibility, the diamonds have been slightly shifted along the x-axis.}
\label{fig:depth_scaling_all}
\end{figure*}

The optimal QAOA depth $p_{\text{opt}}$ can now be derived by minimizing $D$.
More precisely, we minimize the estimated value of $D$ based on the above extrapolation procedure, i.e., we set $P_\text{opt}=P_\text{opt,extr}$ in Eq.~(\ref{eq:Nshots}).
We consider a logarithmically spaced grid of $p$-values
$p\in \{1,2,3\} \cup \{\lfloor 2^{i/4}\rfloor,~i=9,10,11,\dots\}$
and, starting from an initial value $p_0$ (for which we take the optimal value found for the previously calculated instance, or $p_0= 6$ in case of the first instance corresponding to the smallest problem size $N=12$), we increase or decrease $p$ on this grid until a local minimum of $D$ is found. 

\begin{figure*}
\centering
\includegraphics[width=16cm]
{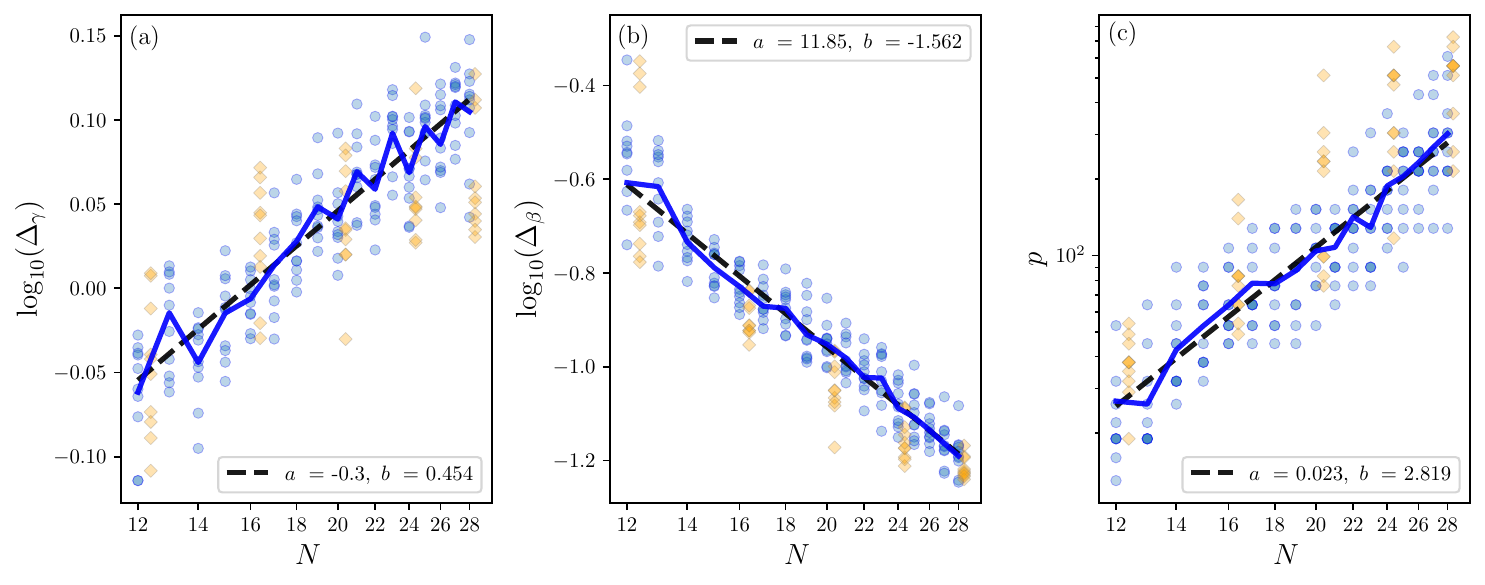}
\caption{Portfolio Optimization: 
optimized linear-ramp parameters (a) $\Delta_{\gamma}$, (b) $\Delta_{\beta}$, and (c) QAOA depth $p$ as a function of number of qubits $N$, depicted as a log-log plot, for the same instances as in Fig.~\ref{fig:depth_scaling_all}(a). The transparent blue dots represent values obtained from extrapolation and the blue lines show the $1/D$-weighted geometric means (see main text) over all instances.
The values for $\Delta_{\gamma}$ as well as for $p$ are increasing with increasing $N$ while the values for $\Delta_{\beta}$ are decreasing. The yellow diamonds represent the optimal values found by a minimizer (as in Fig.~\ref{fig:depth_scaling_all}(a)) and have been shifted along the x-axis for better visibility. 
The black dashed lines display algebraic fits $f(N)=a\cdot N^b$ to the $1/D$-weighted geometric means and determine the universal parameter scaling used in Fig.~\ref{fig:depth_scaling_universal_all}(a).}
\label{fig:Portfolio_parameters}
\end{figure*}

\section{Results and Discussion\label{sec:Results_and_Discussion}}

In this section, we analyze the LR-QAOA runtime scaling for the three use cases (portfolio optimization, feature selection and clustering) introduced in Sec.~\ref{sec:COP}. We increase the problem size from $N=12$ to $N=28$. For each $N$, 10 randomly chosen problem instances are generated, and LR-QAOA circuit parameters ($\Delta_\gamma$, $\Delta_\beta$ and $p$) are found using the extrapolation procedure described in Sect.~\ref{sec:LR-QAOA}. Finally, the
total depth $D$ -- which is proportional to the quantum runtime for finding the optimal solution (see Sec.~\ref{sec:evaluation}) -- is determined  by Eqs.~(\ref{eq:QAOA_circuit},\ref{eq:Popt},\ref{eq:D}) and (\ref{eq:Nshots}) (using a noiseless quantum emulator).
Note that the last step is the only one which involves the execution of a quantum circuit with larger size ($N>10$). Finally, we determine the runtime scaling with increasing $N$ and compare it with the classical result (MQLib/Burer2002) of Sec.~\ref{sec:classical_optimization}.

\subsection{LR-QAOA runtime scaling}

For the case of Portfolio Optimization, the results for the individual 10 random instances per $N$ are displayed by the blue circles in 
Fig.~\ref{fig:depth_scaling_all}(a).
To extract the scaling behavior, we perform an exponential fit, $D \propto 2^{\alpha N}$ (red dashed line).
With this, we find a scaling exponent for the runtime of $\alpha \approx 0.219$, which is smaller than the classical scaling exponent of $\alpha \approx 0.323$
(see Fig.~\ref{Portfolio_classical} bottom right).
This indicates a potential quantum advantage regarding the runtime scaling.

In order to  test our extrapolation procedure,
we further optimized the circuit parameters ($\Delta_\gamma$, $\Delta_\beta$ and $p$) -- starting from the extrapolated parameters
as initial points -- by directly minimizing the actual total depth $D$ (instead of its estimate based on extrapolation).
Since this involves many executions of large quantum circuits (up to $N=28$) for each instance, we restricted ourselves to every fourth problem size $N=12,16,\dots$. All in all, we see that the depths $D$ achieved with the extrapolated parameters (blue circles) are remarkably close to the actual minima (yellow diamonds) in Fig.~\ref{fig:depth_scaling_all}(a). This underlines the effectiveness of the extrapolation scheme.

Fig.~\ref{fig:depth_scaling_all}(b) shows the results for the Feature Selection problem, which we evaluate on the same footing as Portfolio Optimization. Here, we find a scaling for the quantum runtime of $\approx 2^{0.378 N}$. In comparison, the classical runtime exhibits a scaling of $\approx 2^{0.197 N}$ (see Fig.~\ref{Feature_classical}, bottom right) indicating that this combinatorial optimization problem is faster to solve classically. 

A similar picture is observed for the Clustering problem (Moons), as shown in Fig.~\ref{fig:depth_scaling_all}(c) and the MaxCut problem, see Fig.~\ref{fig:depth_scaling_all}(d). In these case, we observe a quantum runtime scaling of $\approx 2^{0.248 N}$ for clustering and $\approx 2^{0.235 N}$ for MaxCut. In comparison, the classical heuristic MQLib/Burer2002 exhibits a scaling of $\approx 2^{0.051 N}$ for clustering (see Fig.~\ref{Moons_classical}, bottom right)
and $\approx 2^{0.156 N}$ for MaxCut (see Fig.~\ref{maxcut_classical}, bottom right)
indicating that the quantum algorithm takes significantly longer to solve the problem. For Clustering, we also observe some outliers with extremely large values of $D$ (e.g., three instances with $D\approx 10^{6}$ at $N=24$ and $28$). 
This suggests that, especially for the Clustering problem, the extrapolation scheme is less stable than, e.g., in the case of Portfolio Optimization and fails in providing good circuit parameters for some instances. This is also confirmed by the results of the direct optimization (yellow diamonds) in Fig.~\ref{fig:depth_scaling_all}(c), which do not exhibit such extreme outliers.

\subsection{Universal parameter scaling}

\begin{figure*}
\centering
\includegraphics[width=16cm]
{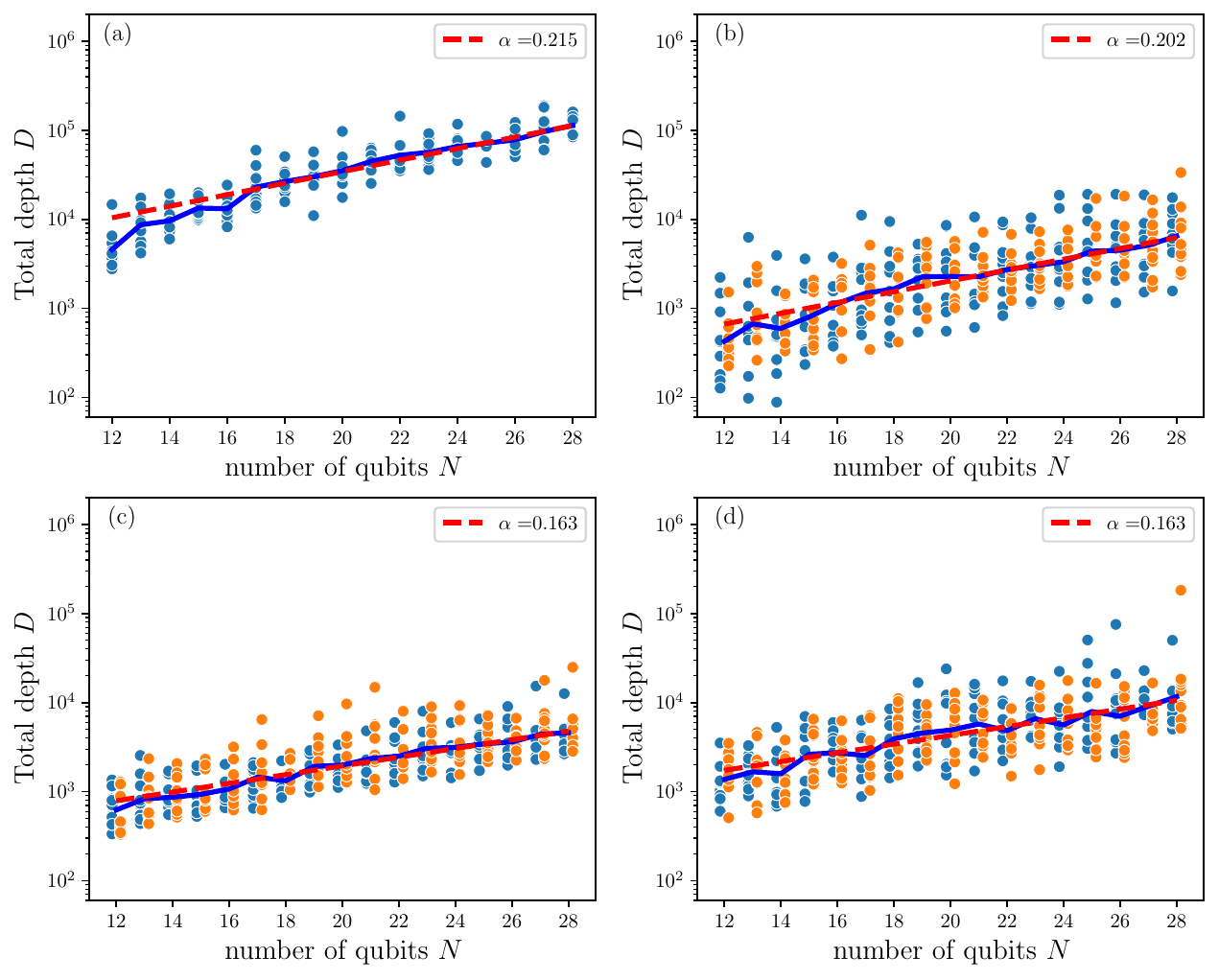}
\caption{Scaling of quantum runtime expressed as total depth $D$, see Eq.~(\ref{eq:D}), with increasing problem size $N$ (number of qubits) using linear-ramp QAOA with universal circuit parameters (defined by the algebraic fits in Figs.~\ref{fig:Portfolio_parameters} and  \ref{fig:Featuresel_parameters}-\ref{fig:Maxcut_parameters}, black dashed lines).
The blue circles refer to the same 10 problem instances per $N$ as in Fig.~\ref{fig:Clustering_parameters}, whereas the orange circles in (b-d) correspond to 10 new randomly chosen problem instances per $N$. As in Fig.~\ref{fig:depth_scaling_all}, the
red dashed lines represent exponential fits with slope $\alpha$ (i.e. $D\propto 2^{\alpha N}$) of the geometric means (blue lines) over all problem
instances per $N$.  
(a) Portfolio Optimization. The result is almost identical to the one in Fig.~\ref{fig:depth_scaling_all}, exhibiting a scaling advantage compared to the classical benchmark
($\alpha=0.22$ vs. $\alpha_{\rm cl}=0.323$).
For the remaining three problem classes (b) Feature Selection, (c) Clustering (Moons) and (d) fully connected weighted random MaxCut, 
the choice of universal parameters significantly improves the LR-QAOA scaling coefficients as compared to Fig.~\ref{fig:depth_scaling_all}. For Feature Selection and MaxCut (b,d), equality with the classical benchmark ($\alpha_{\rm cl}=0.197$ and $0.156$, respectively) is achieved. For Clustering (Moons), the classical benchmark ($\alpha_{\rm cl}=0.051$) is not reached.}
\label{fig:depth_scaling_universal_all}
\end{figure*}

\begin{figure*}
\centering
\includegraphics[width=16cm]
{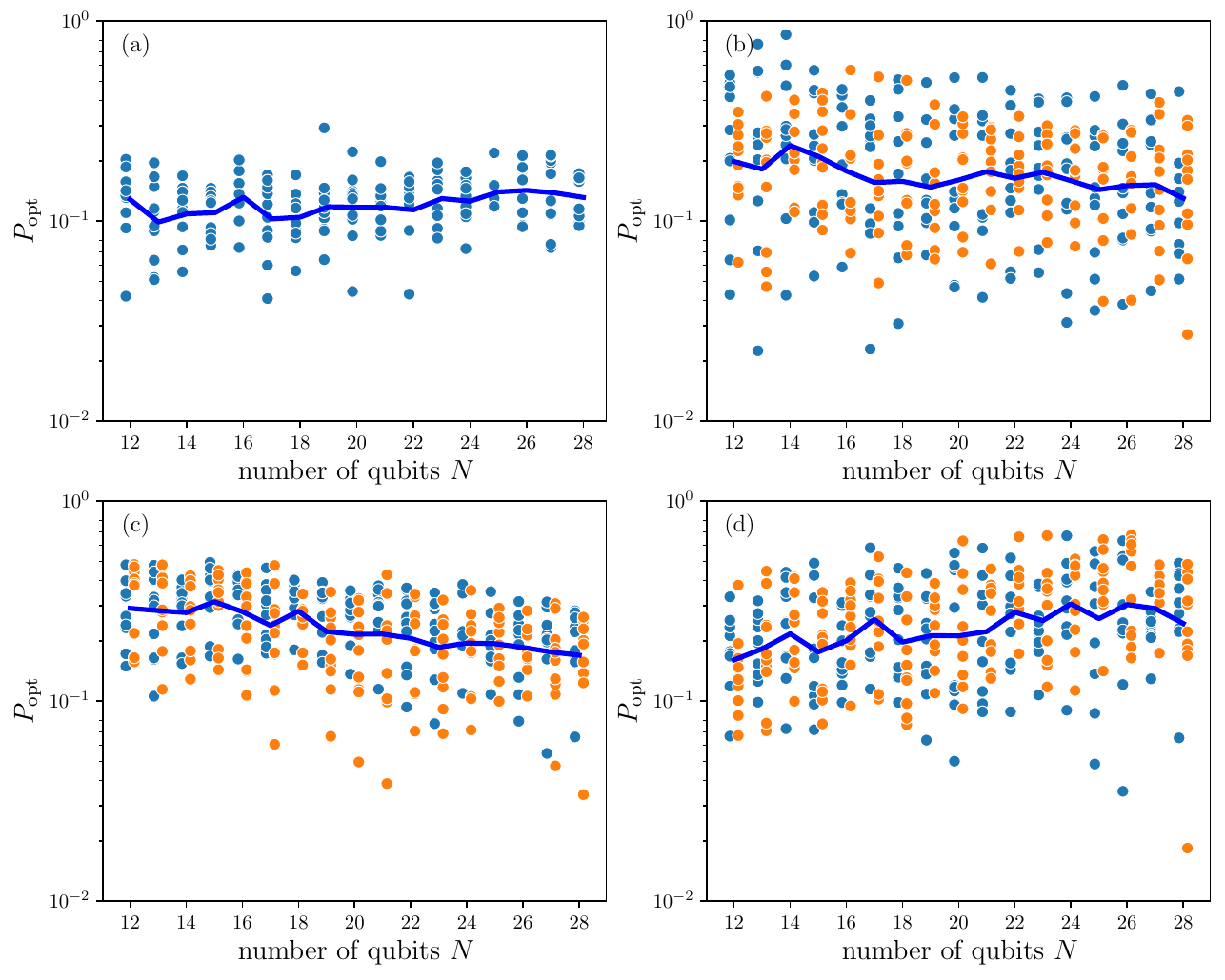}
\caption{Probability $P_{\rm opt}$ to measure the optimal solution after a single run of the LR-QAOA circuit for the same instances and the same universal parameters as in Fig.~\ref{fig:depth_scaling_universal_all} (blue and orange circles). The (geometric) averages $\langle P_{\rm opt}\rangle$ over all problem instances are displayed by the blue lines. On average, the probabilities $\langle P_{\rm opt}\rangle$ remain above $0.1$ and appear to be constant as a function of $N$ in case of (a) Portfolio Optimization and (d) fully connected randomly weighted MaxCut, whereas, for Feature Selection (b) and Clustering (c), a slight decrease of $\langle P_{\rm opt}\rangle$ with increasing $N$ is observed.
}
\label{fig:P_opt_all}
\end{figure*}

In order to reduce these outliers,
we additionally implemented an approach similar to the one proposed in \cite{montanezbarrera2024universalqaoaprotocolevidence}: rather than trying to find LR-QAOA circuit parameters individually for each single instance, the idea is to use the same "universal" parameters for different instances.

Fig.~\ref{fig:Portfolio_parameters} shows the circuit parameters obtained from our extrapolation procedure (blue circles) for the case of Portfolio Optimization.
To determine the universal parameter scaling, we first evaluate -- for each of the three parameters 
$y=\log_{10}(\Delta_\gamma),\log_{10}(\Delta_\beta)$ or $\log_{10}(p)$, respectively --
the $1/D$-weighted average (blue lines), i.e.,
$\langle y\rangle=\left(\sum_{i} \frac{y_i}{D_i}\right)/\left(
\sum_i \frac{1}{D_i}\right)$, where the sum over $i$ refers to the 10 instances per $N$, and the weight $1/D_i$ is employed to suppress those cases where the extrapolation fails in providing good circuit parameters (leading to a large total depth $D_i$ for instance $i$). Remarkably, we see that these averages closely follow algebraic fits using the function $f(N)=a\cdot N^b$ (with $f=\Delta_\gamma$, $\Delta_\beta$ or $p$, respectively), see the dashed lines in Fig.~\ref{fig:Portfolio_parameters}.

These algebraic fits are now taken to define our universal parameter scaling. In other words, we again simulate all instances shown in Fig.~\ref{fig:depth_scaling_all}(a) -- but now using LR-QAOA circuit parameters given by the black dashed line instead of the blue points displayed in Fig.~\ref{fig:Portfolio_parameters}(a). For Portfolio Optimization, the result shown in Fig.~\ref{fig:depth_scaling_universal_all}(a) is similar to the previous one in Fig.~\ref{fig:depth_scaling_all}(a). In particular, the rate of the exponential increase $\simeq 2^{0.215 N}$ is almost unchanged.

For the other three problem classes, however, the results in Fig.~\ref{fig:depth_scaling_universal_all}(b-d)
reveal clear improvements when using the universal parameter scaling (displayed by the black dashed lines in Figs.~\ref{fig:Featuresel_parameters}-\ref{fig:Maxcut_parameters}, see Appendix~\ref{param_plots}). In contrast to Fig.~\ref{fig:depth_scaling_all}(b-d), strong outliers exhibiting extremely large depths $D$ are no longer present. Moreover, we generated, for each problem size $N$, ten additional random instances (orange circles) and evaluated them using the same universal parameter scaling as for the original ten instances (blue circles), with similar results. The scaling coefficients $\alpha$ (determined from all 20 instances per $N$) are significantly smaller than in Fig.~\ref{fig:depth_scaling_all} before. For Feature Selection and the fully connected weighted random MaxCut problem, see Fig.~\ref{fig:depth_scaling_universal_all}(b) and (d), almost equality with the classical benchmarks ($\alpha_{\rm cl}=0.197$ and $0.156$, respectively) is now achieved, whereas, for Clustering (Moons), the classical scaling ($\alpha_{\rm cl}=0.051$) is still better. All in all, we observe that, for our four problem classes, the scaling coefficients using LR-QAOA with universal parameter scaling are remarkably similar to each other (ranging between $\alpha=0.16$ and $0.22$), whereas the classical benchmarks exhibit larger variations (between $\alpha_{\rm cl}=0.051$ and $0.323$).

Concerning the weighted MaxCut problem, a more advantageous scaling coefficient ($\alpha = 0.107$) has been observed in \cite{montanezbarrera2024universalqaoaprotocolevidence}. As we point out in Appendix~\ref{sec:hard_maxcut}, this difference can be traced back to the selection of classically "hard" problem instances performed in \cite{montanezbarrera2024universalqaoaprotocolevidence}.

Finally, we display in Fig.~\ref{fig:P_opt_all} the probabilities $P_{\rm opt}$ of measuring the optimal solution for the same instances and same LR-QAOA parameter scaling as in Fig.~\ref{fig:depth_scaling_universal_all}. In all cases, the probabilities are, on average, above $0.1$. For Portfolio Optimization and MaxCut (a,d), the probability appears (on average) to remain constant (or even slightly increase in case of MaxCut) as a function of the problem size $N$. For Feature Selection and Clustering (Moons), we observe a slight decrease. We note that the behaviour of $P_{\rm opt}$ with increasing $N$ is relevant for the following reason: Since, according to our universal parameter scaling, the QAOA depth $p(N)=a\cdot N^b$ scales algebraically and the depth $d$ of the QAOA circuit is proportional to $p\cdot N$, constant median $P_{\rm opt}$ would imply algebraic, i.e. sub-exponential, scaling of the  median quantum runtime as quantified by the total depth $D$, see Eq.~(\ref{eq:D}).

\section{Conclusion\label{sec:Conclusion}}

In this work, we present an extrapolation method to optimize linear-ramp QAOA parameters, 
enabling us to discard the need for variational optimization employed in standard QAOA. 
We start with the introduction of four
different combinatorial optimization use cases: Portfolio Optimization, Feature Selection, Clustering and weighted MaxCut, which are subsequently solved using classical methods (CPLEX, Gurobi, MQLib Burer2002, Goemans-Williamson) to assess the classical runtime. The scaling of the runtime is investigated by fitting the runtime with an exponential model, which serves as a benchmark for the runtimes of our quantum algorithm. 
In particular, we concentrate on the runtime scaling of the MQLib Burer2002 heuristics, since it exhibits the lowest absolute values of the runtime together with a well-defined scaling behavior over a large range of problem sizes.

Our quantum algorithm is based on linear-ramp QAOA, for which the circuit parameters are determined by a linear ansatz. This ansatz is characterized by three parameters ($\Delta_{\gamma}$, $\Delta_{\beta}$ and depth $p$), which have to be optimized. In this paper, we expose a method to derive those optimized parameters through extrapolation from smaller to larger problem sizes. 
In particular, the extrapolation is based on optimizations of QAOA circuits with reduced size
($\tilde{N}=4,6,8,10$) corresponding to randomly sampled sub-instances of the original problem instance (with size $N$ up to 28). The quantum runtime  is then expressed in terms of the total depth $D=d N_\text{shots}$, i.e., the depth $d$ of a single linear-ramp QAOA circuit times the (median of) the number of shots required to measure the optimal solution on a noiseless quantum emulator.
Comparing quantum and classical runtime scalings, we find an indication of quantum advantage for one of the three use cases mentioned above, i.e., for Portfolio Optimization.

We furthermore showed that, based on the extrapolated optimized circuit parameters, a simple ansatz can be derived, according to which these parameters can be scaled algebraically with increasing problem size $N$, independently of the particular problem instance at hand. Employing this "universal" parameter scaling, leaves the advantageous results in case of Portfolio Optimization unchanged, but significantly improves the scaling of quantum runtime in case of the three other problem classes (achieving equality with the classical benchmark in case of Feature Selection and MaxCut).

One might argue that the cost for finding the optimized circuit parameters should also be taken into account in the evaluation of the quantum runtime scaling. Although, as mentioned above, our extrapolation scheme is based on the analysis of small sub-problems with fixed size (up to $\tilde{N}=10$, independently of $N$),
the computational time required for this analysis nevertheless increases with increasing problem size $N$, due to the fact that the optimal values of the QAOA depth $p$ tend to increase for larger $N$. However, the idea of the present paper is rather to analyze the general potential of LR-QAOA, assuming that suitable circuit parameters are already known or can be efficiently found. An important insight provided by our article is that methods based on extrapolation and scaling of circuit parameters with increasing problem size appear to be a promising tool for this purpose,
and the results presented here may inspire further development of such methods.

Probably the most important open topic concerns the question whether quantum optimization admits  polynomial-time solutions of some instances of combinatorial optimization problems which, classically, can be solved only in exponential time. 
This possibility is suggested by our observation that, for Portfolio Optimization and MaxCut, the median probability $P_{\rm opt}$ of finding the optimal solution after a single execution of the LR-QAOA circuit does not decrease with increasing $N$ while, at the same time, the depth of the circuit scales algebraically. If true
also for larger $N$, this would imply polynomial scaling of the median quantum runtime, i.e., at least half of all problem instances in the respective ensemble could be solved efficiently.

To answer this question, a larger range of problem sizes 
(beyond $N=28$ as in the present paper)
must be addressed in future work, which will either require extensive quantum simulations using high-performance computers and advanced simulation techniques or real, fault-tolerant quantum hardware with sufficient number of qubits (about 50 to 100 or more). 

Furthermore, a more elaborate statistical analysis, not concentrating on the median runtime as we did, but also considering the scaling of different percentiles \cite{ronnow14}, would provide additional insight. Finally, a deeper understanding of the structures of combinatorial optimization problems which determine the hardness of their solution by classical or quantum methods constitutes another challenging goal for future research.

\section*{Acknowledgment} 
\noindent This work is funded by the Ministry of Economic Affairs, Labour and Tourism Baden-Württemberg in the frame of the Competence Center Quantum Computing Baden-Württemberg (projects QORA II and KQCBW24)
and by the European Union’s HORIZON Europe program via project SPINUS (No. 101135699).
A special thanks goes to Michael Krebsbach for his valuable support.  

\section*{Data availability} 
\noindent The data describing the problem instances analyzed in this article are openly available \cite{dataset}

\onecolumngrid
\appendix

\section{Additional classical results}\label{class_data}

\begin{figure*}[h]
\centering
\includegraphics[width=16cm, height=8.0cm]{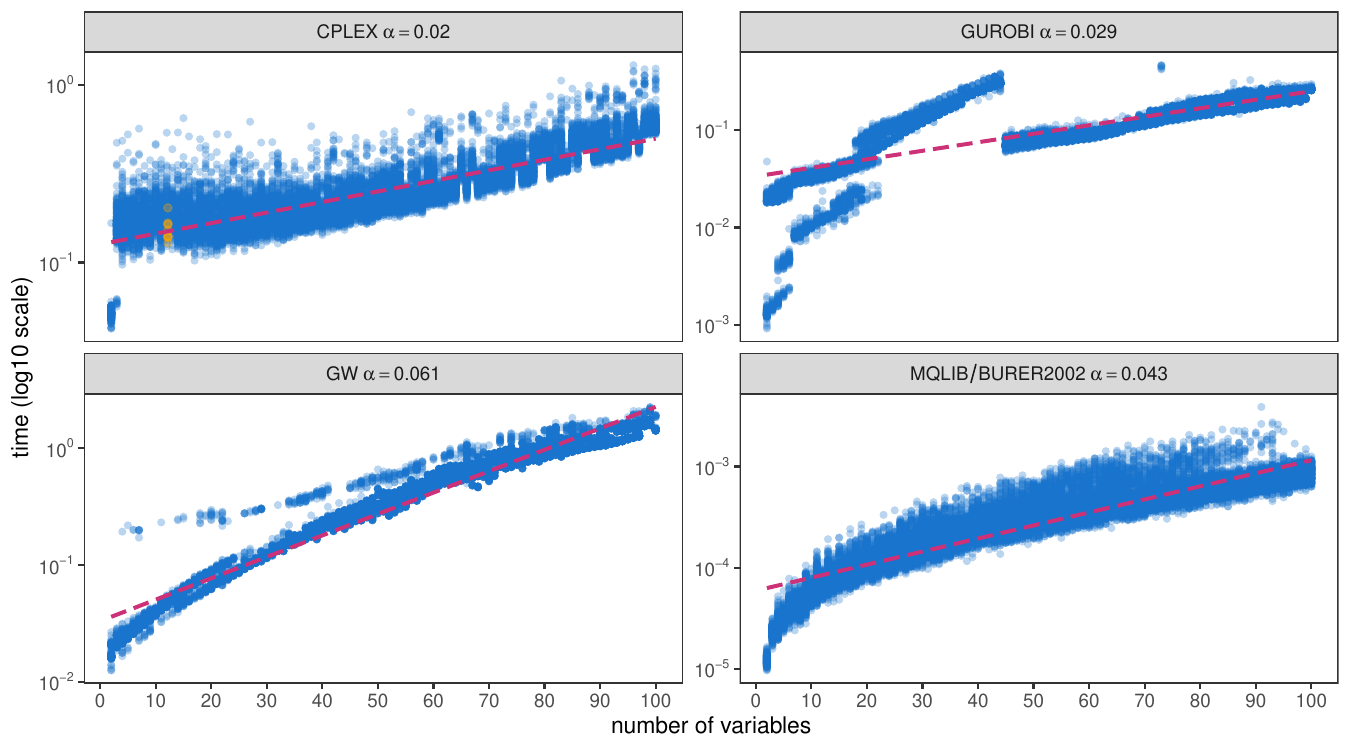}
\caption{Clustering (Blobs): Runtime $T$ (seconds) obtained with CPLEX, Gurobi, Goemans-Williamson and MQLib/Burer2002, as a function of the number $n$ of variables (bits). The dots represent the results 
for the ten problem instances,
with blue indicating it was optimally solved, while orange indicates a non-optimal solution. The slope $\alpha$ of a robust log2-linear model, i.e., $T\propto 2^{\alpha n}$ (red dashed line) is denoted in the header of each figure.}
\label{Blobs_classical}
\end{figure*}

\noindent Fig. \ref{Blobs_classical} illustrates the runtime for the clustering problem using the Blobs dataset. Employing a different dataset results in slightly varying outcomes. This emphasizes the influence of both the dataset and the selected instances on runtime scaling. Notably, the Goemans-Wiliamson algorithm is able to find the optimal solution in this case, and exhibits a better scaling factor $\alpha$ (slope of exponential fit) compared to the Moons dataset.  

\section{LR-QAOA parameters for feature selection and clustering}\label{param_plots}

\begin{figure*}[ht!] 
\centering
\includegraphics[width=0.95\textwidth]{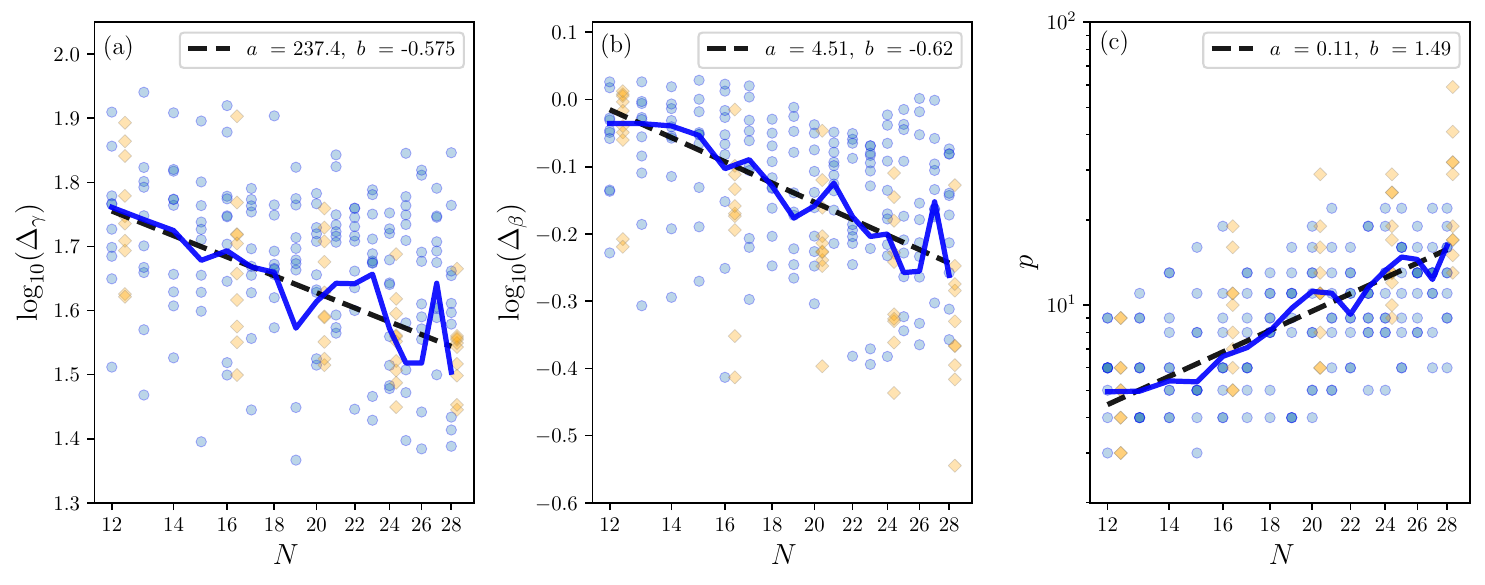}
\caption{Feature Selection: same representation as for portfolio optimization of optimized linear-ramp parameters (a): $\Delta_{\gamma}$ , (b): $\Delta_{\beta}$, and (c): QAOA depth $p$ as a function of number of qubits $N$ on a log-log scale. In this case, the logarithmic values for $\Delta_{\gamma}$ and for $\Delta_{\beta}$ are relatively constant with $N$. For the depth $p$, we observe a nearly linear scaling with respect to $N$. The yellow diamonds again represent the optima found by a minimizer.
The black dashed lines display algebraic fits $f(N)=a\cdot N^b$ to the 
$1/D$-weighted geometric means (blue lines) 
and determine the universal parameter scaling used in Fig.~\ref{fig:depth_scaling_universal_all}(b).
}
\label{fig:Featuresel_parameters}
\end{figure*}

\noindent Fig. \ref{fig:Featuresel_parameters} shows the same parameters for the feature selection problem as shown in the main text (Fig.~\ref{fig:Portfolio_parameters}) for portfolio optimization.  In this case, $\Delta_{\gamma}$ and $\Delta_{\beta}$ remain relatively constant as the problem size increases.

\begin{figure*}[ht!] 
\centering
\includegraphics[width=0.95\textwidth]{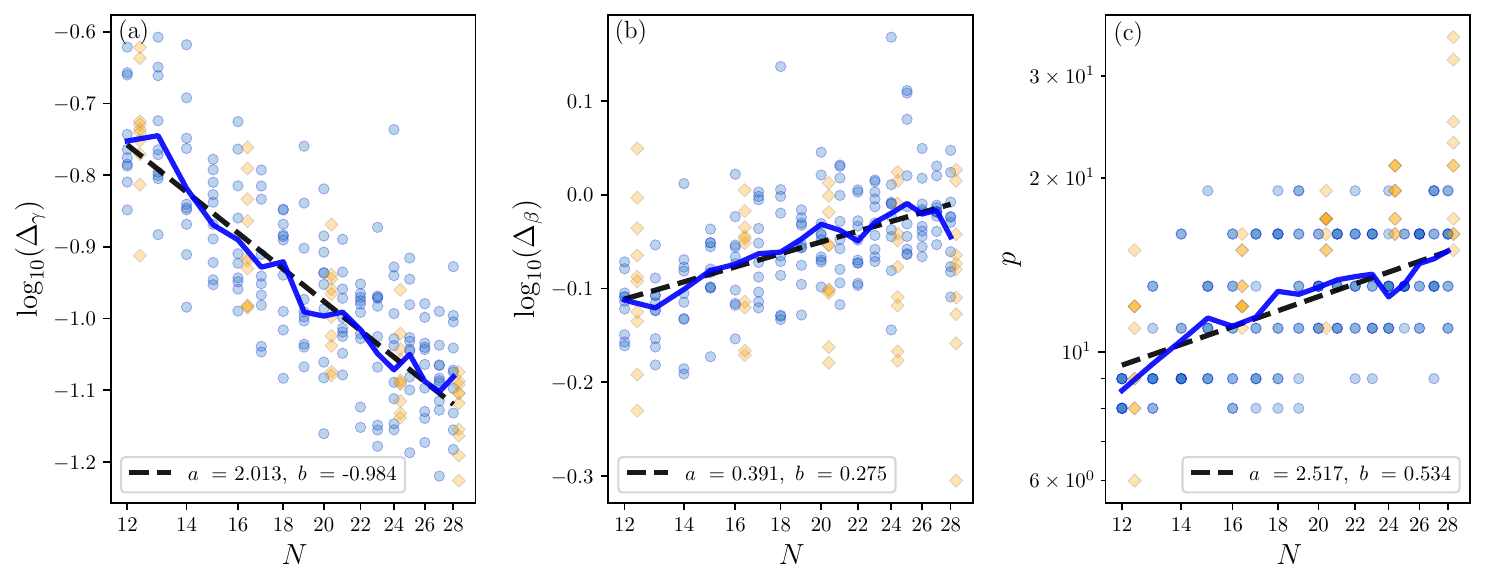}
\caption{Clustering (Moons): optimized linear-ramp parameters (a): $\Delta_{\gamma}$ , (b): $\Delta_{\beta}$, and (c): QAOA depth $p$ as a function of number of qubits $N$ on a log-log scale. In this case, the values for $\Delta_{\gamma}$ are decreasing while the values for $\Delta_{\beta}$ are slightly increasing with $N$. For the depth $p$, we observe a scaling of $\propto \sqrt{N}$. The optima found by a minimizer are represented by the yellow diamonds. The black dashed lines display algebraic fits $f(N)=a\cdot N^b$ to the 
$1/D$-weighted geometric means (blue lines) and determine the universal parameter scaling used in Fig.~\ref{fig:depth_scaling_universal_all}(c).}
\label{fig:Clustering_parameters}
\end{figure*}

The parameters for the clustering problem are illustrated in Fig. \ref{fig:Clustering_parameters}, again showing a different trend compared to the other problems. Here, we see $\Delta_{\gamma}$ decreasing and $\Delta_{\beta}$ increasing. 

\begin{figure*}[ht!] 
\centering
\includegraphics[width=0.95\textwidth]{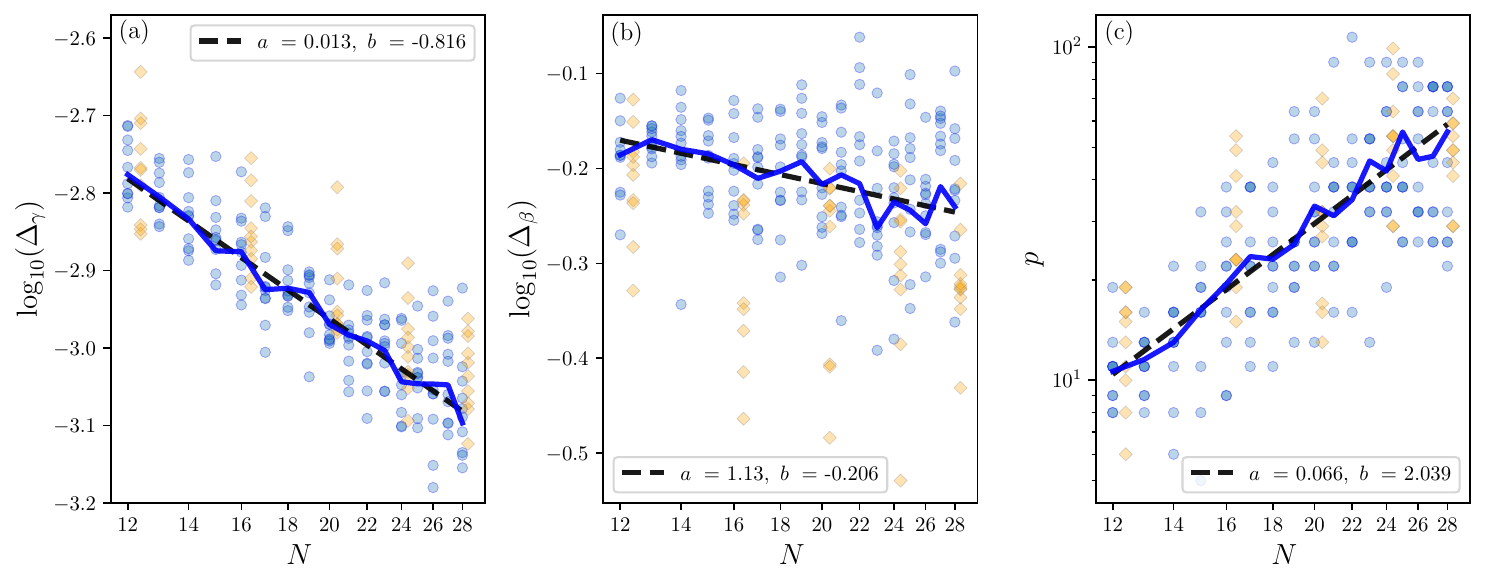}
\caption{MaxCut: optimized linear-ramp parameters (a) $\Delta_{\gamma}$, (b) $\Delta_{\beta}$, and (c) QAOA depth $p$ as a function of number of qubits $N$ on a log-log scale. The black dashed lines display algebraic fits $f(N)=a\cdot N^b$ to the $1/D$-weighted geometric means (blue lines) and determine the universal parameter scaling used in Fig.~\ref{fig:depth_scaling_universal_all}(c).}
\label{fig:Maxcut_parameters}
\end{figure*}

\section{Distribution of the QUBO-Data}\label{boxplots_data}

All optimization problems considered above are defined by their corresponding QUBO matrices. To compare them, we performed the following analysis: For a given problem size (here, $N=20$) and for each problem, we calculated the maximum absolute value across all instances and normalized the respective QUBO matrix by this value.
Subsequently, we computed the box plots for both the diagonal and non-diagonal elements. In Fig.~(\ref{BoxPlot_all}), we compare the box plots of the portfolio optimization problem with those of the other problems: The distributions of both the diagonal and non-diagonal elements of the portfolio optimization QUBO are much narrower than those of the other problems. This is due to the fact that portfolio optimization includes a penalty term to enforce the constraint, i.e., the maximum number of assets allowed in the portfolio.

\begin{figure*}[h]
\centering
\includegraphics[width=16cm, height=8cm]
{combined_boxplots_Max-Cut_Moons_Blobs_Feature-Selection_Portfolio-Optimization.png}
\caption{Comparison of box plots representing the distribution of QUBO matrix elements between all problem types under consideration for problem size $N=20$. The distributions of the QUBO matrices' diagonal elements are shown in the left figure; the distribution of the non-diagonal elements in the right figure. The QUBO matrices have been normalized by the following factors:
$7.39\times 10^{-5}$ (MaxCut), $0.0102$ (Clustering/Moons), $0.0112$ (Clustering/Blobs), $3.45$ (Feature Selection), and $0.0215$ (Portfolio Optimization).
}
\label{BoxPlot_all}
\end{figure*}

\section{Results for hard MaxCut instances}
\label{sec:hard_maxcut}

\begin{figure*}
\centering
\includegraphics[width=10cm]
{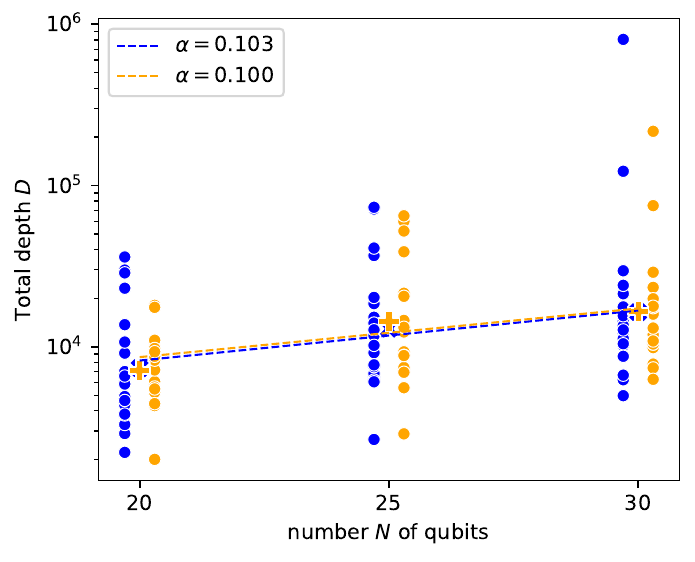}
\caption{Scaling of quantum runtime expressed as total depth $D$, see Eq.~(\ref{eq:D}), for the set of selected, classically hard weighted MaxCut instances ($N=20,25,30$) considered in \cite{montanezbarrera2024universalqaoaprotocolevidence}. The blue dots (slightly shifted to the left for better visibility) show the depths obtained with the LR-QAOA circuit parameters used in \cite{montanezbarrera2024universalqaoaprotocolevidence}, whereas the orange dots (shifted to the right) display the results of our universal parameter scaling ansatz. The blue and orange crosses indicate the (geometric) mean values, respectively. The dashed lines represent exponential fits $D\propto 2^{\alpha N}$, with almost identical scaling coefficients ($\alpha=0.103$ and $0.100$) -- similar to the one ($\alpha=0.107$) found in \cite{montanezbarrera2024universalqaoaprotocolevidence} for problem sizes up to $N=40$.}
\label{fig:depth_scaling_maxcut_hard_instances}
\end{figure*}

Concerning the weighted random MaxCut problem, our result ($\alpha=0.163$) for the scaling coefficient, see Fig.~\ref{fig:depth_scaling_universal_all}(d), differs from the one ($\alpha=0.107$) obtained in \cite{montanezbarrera2024universalqaoaprotocolevidence}. The latter result has been obtained for a selected set of instances ($N=20,25,30,35$ and $40$) exhibiting the longest classical runtimes. In this appendix, we present results for the same instances (up to $N=30$) using our universal parameter scaling obtained from Fig.~\ref{fig:Maxcut_parameters}. 
Note, however, that the normalization of QUBO matrices (such that $\max_{i\neq j} Q_{ij}=2$) employed in \cite{montanezbarrera2024universalqaoaprotocolevidence} leads to a corresponding shift of $\Delta_\gamma$ values for each instance, see the blue dots in Fig.~\ref{fig:Maxcut_parameters}, which changes the fit parameters to $a=5.945$ and $b=-0.800$ with respect to the values displayed in Fig.~\ref{fig:Maxcut_parameters}(a).

Thereby, we obtain the following LR-QAOA parameters:
$\Delta_\gamma=0.542$, $\Delta_\beta=0.608$, $p=30$ for $N=20$, $\Delta_\gamma=0.453$, $\Delta_\beta=0.581$, $p=47$ for $N=25$ and
$\Delta_\gamma=0.392$, $\Delta_\beta=0.560$, $p=67$ for $N=30$. In contrast, the values $\Delta_\gamma=0.4$, $\Delta_\beta=0.4$, $p=100$ for $N=20$, $\Delta_\gamma=0.5$, $\Delta_\beta=0.4$, $p=50$ for $N=25$ and
$\Delta_\gamma=0.4$, $\Delta_\beta=0.5$, $p=100$ for $N=30$ are used in \cite{montanezbarrera2024universalqaoaprotocolevidence}. In Fig.~\ref{fig:depth_scaling_maxcut_hard_instances}, we see that these two sets of parameters achieve (at least on average) almost identical results, with scaling coefficients $\alpha=0.103$ and $0.100$, respectively.

Comparing the values of $D$ in Figs.~\ref{fig:depth_scaling_universal_all}(d) and
Fig.~\ref{fig:depth_scaling_maxcut_hard_instances}, we see that the instances, which have been selected due to their long classical runtimes, are also more difficult to solve for LR-QAOA, leading to larger $D$. Moreover, this increase of $D$ appears to be more strongly pronounced for smaller problem sizes, which, in total amounts to a smaller scaling coefficient $\alpha$.

\newpage
\twocolumngrid
\bibliography{bibliography}

@article{Brandhofer2022,
  title = {Benchmarking the performance of portfolio optimization with {QAOA}},
  author = {Brandhofer, Sebastian and Braun, Daniel and Dehn, Vanessa and Hellstern, Gerhard and Hüls, Matthias and Ji, Yanjun and Polian, Ilia and Bhatia, Amandeep Singh and Wellens, Thomas},
  journal = {Quantum Information Processing},
  volume = {22},
  issue = {1},
  pages = {25},
  numpages = {27},
  year = {2022},
  month = {Dec},
  publisher = {Springer},
  doi = {10.1007/s11128-022-03766-5 },
  url = {https://doi.org/10.1007/s11128-022-03766-5}                                              
}

@article{Goemans1995, 
author = {Goemans, Michel X. and Williamson, David P.},
title = {Improved Approximation Algorithms for Maximum Cut and Satisfiability Problems Using Semidefinite Programming},
year = {1995},
issue_date = {Nov. 1995},
publisher = {Association for Computing Machinery},
address = {New York, NY, USA},
volume = {42},
number = {6},
issn = {0004-5411},
url = {https://doi.org/10.1145/227683.227684                                        }       ,
doi = {10.1145/227683.227684},
journal = {J. ACM},
month = {nov},
pages = {1115–1145},
numpages = {31}
}

@article{zhou2020,
  title = {Quantum Approximate Optimization Algorithm: Performance, Mechanism, and Implementation on Near-Term Devices},
  author = {Zhou, Leo and Wang, Sheng-Tao and Choi, Soonwon and Pichler, Hannes and Lukin, Mikhail D.},
  journal = {Phys. Rev. X},
  volume = {10},
  issue = {2},
  pages = {021067},
  numpages = {23},
  year = {2020},
  month = {Jun},
  publisher = {American Physical Society},
  doi = {10.1103/PhysRevX.10.021067  },
  url = {https://link.aps.org/doi/10.1103/PhysRevX.10.021067}                                                     
}

@article{Farhi2022quantumapproximate,
  doi = {10.22331/q-2022-07-07-759  },
  url = {https://doi.org/10.22331/q-2022-07-07-759}                                                           ,
  title = {The {Q}uantum {A}pproximate {O}ptimization {A}lgorithm and the {S}herrington-{K}irkpatrick {M}odel at {I}nfinite {S}ize},
  author = {Farhi, Edward and Goldstone, Jeffrey and Gutmann, Sam and Zhou, Leo},
  journal = {{Quantum}},
  issn = {2521-327X},
  publisher = {{Verein zur F{\"{o}}rderung des Open Access Publizierens in den Quantenwissenschaften}},
  volume = {6},
  pages = {759},
  month = jul,
  year = {2022}
}

@misc{farhi2014quantum,
      title={A Quantum Approximate Optimization Algorithm}, 
      author={Edward Farhi and Jeffrey Goldstone and Sam Gutmann},
      year={2014},
      eprint={1411.4028},
      archivePrefix={arXiv},
      primaryClass={quant-ph}
}

@misc{farhi25,
      title={Lower bounding the MaxCut of high girth 3-regular graphs using the {QAOA}}, 
      author={Edward Farhi and Sam Gutmann and Daniel Ranard and Benjamin Villalonga},
      year={2025},
      eprint={2503.12789},
      archivePrefix={arXiv},
      primaryClass={quant-ph}
}

@article{Grange_2023,
   title={An introduction to variational quantum algorithms for combinatorial optimization problems},
   volume={21},
   ISSN={1614-2411},
   url={http://dx.doi.org/10.1007/s10288-023-00549-1                                                }  ,
   DOI={10.1007/s10288-023-00549-1},
   number={3},
   journal={4OR},
   publisher={Springer Science and Business Media LLC},
   author={Grange, Camille and Poss, Michael and Bourreau, Eric},
   year={2023},
   month=jul, pages={363–403} }

@inproceedings{Shaydulin_2019,
   title={Evaluating Quantum Approximate Optimization Algorithm: A Case Study},
   url={http://dx.doi.org/10.1109/IGSC48788.2019.8957201                                                 } ,
   DOI={10.1109/igsc48788.2019.8957201 },
   booktitle={2019 Tenth International Green and Sustainable Computing Conference (IGSC)},
   publisher={IEEE},
   author={Shaydulin, Ruslan and Alexeev, Yuri},
   year={2019},
   month=oct, pages={1–6} }

@article{BLEKOS20241,
title = {A review on Quantum Approximate Optimization Algorithm and its variants},
journal = {Physics Reports},
volume = {1068},
pages = {1-66},
year = {2024},
issn = {0370-1573},
doi = {https://doi.org/10.1016/j.physrep.2024.03.002} ,
url = {https://www.sciencedirect.com/science/article/pii/S0370157324001078},
author = {Kostas Blekos and Dean Brand and Andrea Ceschini and Chiao-Hui Chou and Rui-Hao Li and Komal Pandya and Alessandro Summer},}

@article{Markowitz1952,
 ISSN = {00221082, 15406261},
 URL = {http://www.jstor.org/stable/2975974},
 author = {Harry Markowitz},
 journal = {The Journal of Finance},
 number = {1},
 pages = {77--91},
 publisher = {[American Finance Association, Wiley]},
 title = {Portfolio Selection},
 urldate = {2024-12-02},
 volume = {7},
 year = {1952}
}

@article{Sureshbabu2024parametersettingin,
  doi = {10.22331/q-2024-01-18-1231},
  url = {https://doi.org/10.22331/q-2024-01-18-1231}                                      ,
  title = {Parameter {S}etting in {Q}uantum {A}pproximate {O}ptimization of {W}eighted {P}roblems},
  author = {Sureshbabu, Shree Hari and Herman, Dylan and Shaydulin, Ruslan and Basso, Joao and Chakrabarti, Shouvanik and Sun, Yue and Pistoia, Marco},
  journal = {{Quantum}},
  issn = {2521-327X},
  publisher = {{Verein zur F{\"{o}}rderung des Open Access Publizierens in den Quantenwissenschaften}},
  volume = {8},
  pages = {1231},
  month = jan,
  year = {2024}
}

@article{PhysRevResearch.6.023171,
  title = {Parameter-setting heuristic for the quantum alternating operator ansatz},
  author = {Sud, James and Hadfield, Stuart and Rieffel, Eleanor and Tubman, Norm and Hogg, Tad},
  journal = {Phys. Rev. Res.},
  volume = {6},
  issue = {2},
  pages = {023171},
  numpages = {14},
  year = {2024},
  month = {May},
  publisher = {American Physical Society},
  doi = {10.1103/PhysRevResearch.6.023171},
  url = {https://link.aps.org/doi/10.1103/PhysRevResearch.6.023171}                                     
}

@misc{kremenetski2021quantumalternatingoperatoransatz,
      title={Quantum Alternating Operator Ansatz {(QAOA)} Phase Diagrams and Applications for Quantum Chemistry}, 
      author={Vladimir Kremenetski and Tad Hogg and Stuart Hadfield and Stephen J. Cotton and Norm M. Tubman},
      year={2021},
      eprint={2108.13056},
      archivePrefix={arXiv},
      primaryClass={quant-ph},
      url={https://arxiv.org/abs/2108.13056}, 
}

@article{Sack2021quantumannealing,
  doi = {10.22331/q-2021-07-01-491},
  url = {https://doi.org/10.22331/q-2021-07-01-491}                                    ,
  title = {Quantum annealing initialization of the quantum approximate optimization algorithm},
  author = {Sack, Stefan H. and Serbyn, Maksym},
  journal = {{Quantum}},
  issn = {2521-327X},
  publisher = {{Verein zur F{\"{o}}rderung des Open Access Publizierens in den Quantenwissenschaften}},
  volume = {5},
  pages = {491},
  month = jul,
  year = {2021}
}

@misc{mbeng2019quantumannealingjourneydigitalization,
      title={Quantum Annealing: a journey through Digitalization, Control, and hybrid Quantum Variational schemes}, 
      author={Glen Bigan Mbeng and Rosario Fazio and Giuseppe Santoro},
      year={2019},
      eprint={1906.08948},
      archivePrefix={arXiv},
      primaryClass={quant-ph},
      url={https://arxiv.org/abs/1906.08948}, 
}

@misc{crooks2018performancequantumapproximateoptimization,
      title={Performance of the Quantum Approximate Optimization Algorithm on the Maximum Cut Problem}, 
      author={Gavin E. Crooks},
      year={2018},
      eprint={1811.08419},
      archivePrefix={arXiv},
      primaryClass={quant-ph},
      url={https://arxiv.org/abs/1811.08419}, 
}

@misc{kremenetski2023quantumalternatingoperatoransatz,
      title={Quantum Alternating Operator Ansatz {(QAOA)} beyond low depth with gradually changing unitaries}, 
      author={Vladimir Kremenetski and Anuj Apte and Tad Hogg and Stuart Hadfield and Norm M. Tubman},
      year={2023},
      eprint={2305.04455},
      archivePrefix={arXiv},
      primaryClass={quant-ph},
      url={https://arxiv.org/abs/2305.04455}, 
}

@article{dranov98, 
    author = {Dranov, A. and Kellendonk, J. and Seiler, R.},
    title = {Discrete time adiabatic theorems for quantum mechanical systems},
    journal = {Journal of Mathematical Physics},
    volume = {39},
    number = {3},
    pages = {1340-1349},
    year = {1998},
    month = {03},
    issn = {0022-2488},
    doi = {10.1063/1.532382 },
    url = {https://doi.org/10.1063/1.532382                                 }  ,
    eprint = {https://pubs.aip.org/aip/jmp/article-pdf/39/3/1340/19166477/1340\_1\_online.pdf},
}

@article{montanezbarrera2024universalqaoaprotocolevidence,
      title={Towards a universal {QAOA} protocol: Evidence of a scaling advantage in solving some combinatorial optimization problems}, 
      author={J. A. Montanez-Barrera and Kristel Michielsen},
      journal = {npj Quantum Inf.},
      year={2025},
      pages = {131},
      volume = {11},
      url={https://www.nature.com/articles/s41534-025-01082-1}
}

@misc{brandao2018fixedcontrolparametersquantum,
      title={For Fixed Control Parameters the Quantum Approximate Optimization Algorithm's Objective Function Value Concentrates for Typical Instances}, 
      author={Fernando G. S. L. Brandao and Michael Broughton and Edward Farhi and Sam Gutmann and Hartmut Neven},
      year={2018},
      eprint={1812.04170},
      archivePrefix={arXiv},
      primaryClass={quant-ph},
      url={https://arxiv.org/abs/1812.04170}, 
}

@misc{sakai2024linearlysimplifiedqaoaparameters,
      title={Linearly simplified {QAOA} parameters and transferability}, 
      author={Ryo Sakai and Hiromichi Matsuyama and Wai-Hong Tam and Yu Yamashiro and Keisuke Fujii},
      year={2024},
      eprint={2405.00655},
      archivePrefix={arXiv},
      primaryClass={quant-ph},
      url={https://arxiv.org/abs/2405.00655}, 
}

@article{Shaydulin2021,
  title = {Classical symmetries and the Quantum Approximate Optimization Algorithm},
  author = {Ruslan Shaydulin and Stuart Hadfield and Tad Hogg and Ilya Safro},
  journal = {Quantum Information Processing},
  volume = {20},
  issue = {11},
  pages = {359},
  numpages = {28},
  year = {2021},
  month = {Oct},
  publisher = {Springer},
  doi = {10.1007/s11128-021-03298-4 },
  url = {https://doi.org/10.1007/s11128-021-03298-4}                                              
}

@article{Shaydulin2024,
doi={10.1126/sciadv.adm6761},          
author = {Ruslan Shaydulin  and Changhao Li  and Shouvanik Chakrabarti  and Matthew DeCross  and Dylan Herman  and Niraj Kumar  and Jeffrey Larson  and Danylo Lykov  and Pierre Minssen  and Yue Sun  and Yuri Alexeev  and Joan M. Dreiling  and John P. Gaebler  and Thomas M. Gatterman  and Justin A. Gerber  and Kevin Gilmore  and Dan Gresh  and Nathan Hewitt  and Chandler V. Horst  and Shaohan Hu  and Jacob Johansen  and Mitchell Matheny  and Tanner Mengle  and Michael Mills  and Steven A. Moses  and Brian Neyenhuis  and Peter Siegfried  and Romina Yalovetzky  and Marco Pistoia },
title = {Evidence of scaling advantage for the quantum approximate optimization algorithm on a classically intractable problem},
journal = {Science Advances},
volume = {10},
number = {22},
pages = {eadm6761},
year = {2024},
doi = {10.1126/sciadv.adm6761},
}

@article{Saiphet_2021,
doi = {10.1088/1742-6596/1719/1/012100 },
url = {https://dx.doi.org/10.1088/1742-6596/1719/1/012100}                                                                 ,
year = {2021},
month = {jan},
publisher = {IOP Publishing},
volume = {1719},
number = {1},
pages = {012100},
author = {Jirawat Saiphet and Sujin Suwanna and Thiparat Chotibut and Areeya Chantasri},
title = {Quantum approximate optimization and k-means algorithms for data clustering},
journal = {Journal of Physics: Conference Series}
}

@misc{otterbach2017unsupervised,
      title={Unsupervised Machine Learning on a Hybrid Quantum Computer}, 
      author={J. S. Otterbach and R. Manenti and N. Alidoust and A. Bestwick and M. Block and B. Bloom and S. Caldwell and N. Didier and E. Schuyler Fried and S. Hong and P. Karalekas and C. B. Osborn and A. Papageorge and E. C. Peterson and G. Prawiroatmodjo and N. Rubin and Colm A. Ryan and D. Scarabelli and M. Scheer and E. A. Sete and P. Sivarajah and Robert S. Smith and A. Staley and N. Tezak and W. J. Zeng and A. Hudson and Blake R. Johnson and M. Reagor and M. P. da Silva and C. Rigetti},
      year={2017},
      eprint={1712.05771},
      archivePrefix={arXiv},
      primaryClass={quant-ph}
}

@article{cplex2009v12,
  title={V12. 1: User’s Manual for CPLEX},
  author={Cplex, IBM ILOG},
  journal={International Business Machines Corporation},
  volume={46},
  number={53},
  pages={157},
  year={2009}
}

@misc{gurobi,
  author = {{Gurobi Optimization, LLC}},
  title = {{Gurobi Optimizer Reference Manual}},
  year = 2023,
  url = "https://www.gurobi.com"
}

@article{DunningEtAl2018,
  title={What Works Best When? A Systematic Evaluation of Heuristics for Max-Cut and {QUBO}},
  author={Dunning, Iain and Gupta, Swati and Silberholz, John},
  year={2018},
  journal={{INFORMS} Journal on Computing},
  volume={30},
  number={3}
}

@misc{Qiskit,
    author = {{Qiskit contributors}},
    title = {Qiskit: An Open-source Framework for Quantum Computing},
    year = {2023},
    doi = {10.5281/zenodo.2573505}
}

@article{lloyd1982least,
  title={Least squares quantization in PCM},
  author={Lloyd, Stuart},
  journal={IEEE transactions on information theory},
  volume={28},
  number={2},
  pages={129--137},
  year={1982},
  publisher={IEEE}
}

@article{Sibson73,
  author = {Sibson, R.},
  description = {WSD},
  journal = {The Computer Journal},
  number = 1,
  pages = {30--34},
  title = {{SLINK}: an optimally efficient algorithm for the single-link cluster method},
  volume = 16,
  year = 1973
}

@inproceedings{ester1996density,
  title={A density-based algorithm for discovering clusters in large spatial databases with noise.},
  author={Ester, Martin and Kriegel, Hans-Peter and Sander, Jorg and Xu, Xiaowei and others},
  booktitle={kdd},
  volume={96},
  number={34},
  pages={226--231},
  year={1996}
}

@misc{Glover2019,
      title={A Tutorial on Formulating and Using {QUBO} Models}, 
      author={Fred Glover and Gary Kochenberger and Yu Du},
      year={2019},
      eprint={1811.11538},
      archivePrefix={arXiv},
      primaryClass={cs.DS},
      url={https://arxiv.org/abs/1811.11538}, 
}

@article{HellsternDehnZaefferer2024,
author = {Hellstern, Gerhard and Dehn, Vanessa and Zaefferer, Martin},
title = {Quantum computer based feature selection in machine learning},
journal = {IET Quantum Communication},
volume = {5},
number = {3},
pages = {232-252},
doi = {https://doi.org/10.1049/qtc2.12086}  ,
year = {2024}
}

@misc{germancreditdata,
  author       = {Hofmann, Hans},
  title        = {{Statlog (German Credit Data)}},
  year         = {1994},
  howpublished = {UCI Machine Learning Repository},
  note         = {{DOI}: https://doi.org/10.24432/C5NC77      }        
}

@article{scikitlearn,
  title={Scikit-learn: Machine Learning in {P}ython},
  author={Pedregosa, F. and Varoquaux, G. and Gramfort, A. and Michel, V.
          and Thirion, B. and Grisel, O. and Blondel, M. and Prettenhofer, P.
          and Weiss, R. and Dubourg, V. and Vanderplas, J. and Passos, A. and
          Cournapeau, D. and Brucher, M. and Perrot, M. and Duchesnay, E.},
  journal={Journal of Machine Learning Research},
  volume={12},
  pages={2825--2830},
  year={2011}
}

@article{PRXQuantum.5.030348,
  title = {Solving Boolean Satisfiability Problems With The Quantum Approximate Optimization Algorithm},
  author = {Boulebnane, Sami and Montanaro, Ashley},
  journal = {PRX Quantum},
  volume = {5},
  issue = {3},
  pages = {030348},
  numpages = {32},
  year = {2024},
  month = {Sep},
  publisher = {American Physical Society},
  doi = {10.1103/PRXQuantum.5.030348},
  url = {https://link.aps.org/doi/10.1103/PRXQuantum.5.030348}     
}

@article{PhysRevE.99.063314,
  title = {Fair sampling of ground-state configurations of binary optimization problems},
  author = {Zhu, Zheng and Ochoa, Andrew J. and Katzgraber, Helmut G.},
  journal = {Phys. Rev. E},
  volume = {99},
  issue = {6},
  pages = {063314},
  numpages = {6},
  year = {2019},
  month = {Jun},
  publisher = {American Physical Society},
  doi = {10.1103/PhysRevE.99.063314},
  url = {https://link.aps.org/doi/10.1103/PhysRevE.99.063314}  
}

@article{born28,
  title = {Beweis des {A}diabatensatzes},
  author = {Born, M. and Fock, V.},
  journal = {Zeitschrift für Physik},
  volume = {51},
  pages = {165},
  year = {1928},
  doi = {10.1007/BF01343193}
}

@article{ronnow14,
author = {Troels F. Rønnow  and Zhihui Wang  and Joshua Job  and Sergio Boixo  and Sergei V. Isakov  and David Wecker  and John M. Martinis  and Daniel A. Lidar  and Matthias Troyer},
title = {Defining and detecting quantum speedup},
journal = {Science},
volume = {345},
number = {6195},
pages = {420-424},
year = {2014},
doi = {10.1126/science.1252319},
URL = {https://www.science.org/doi/abs/10.1126/science.1252319}  ,
eprint = {https://www.science.org/doi/pdf/10.1126/science.1252319}  ,
}

@misc{dataset,
      title={Extrapolation method to optimize linear-ramp quantum approximate optimization algorithm parameters: dataset}, 
      author={Thomas Wellens and Vanessa Dehn and Gerhard Hellstern and Karthik Jayadevan and Florentin Reiter and Thomas Wellens},
      year={2026},
      journal = {Zenodo},
      doi = {10.5281/zenodo.18302992}
}
\end{document}